\documentclass[aps,notitlepage,nofootinbib,longbibliography,twocolumn,superscriptaddress,10pt]{revtex4-2}

\usepackage{bbold}
\usepackage{simplewick}

\usepackage[usenames,dvipsnames]{color}
\usepackage[colorlinks=true,citecolor=PineGreen,linkcolor=blue]{hyperref}
\usepackage[normalem]{ulem}
\usepackage{amsmath}
\usepackage{amsthm, amssymb}
\usepackage{physics}
\usepackage{enumitem}
\usepackage{slashed}
\usepackage{graphicx}
\usepackage{xcolor}
\usepackage{tikz}
\usetikzlibrary{calc,fadings,decorations.pathreplacing,shapes,shapes.multipart,arrows,shapes.misc,intersections,positioning,patterns}
\usepackage{bm}
\usepackage{bbm}
\usepackage[colorlinks=true,citecolor=PineGreen,linkcolor=blue]{hyperref}
\usepackage{dsfont}
\usepackage{changepage}
\usepackage{array}
\usepackage{soul}
\usepackage[capitalise]{cleveref}
\crefname{section}{Sec.}{Secs.}
\Crefname{section}{Sec.}{Secs.}
\usepackage{xifthen}
\usepackage{xargs}
\usepackage{braket}
\newcommand{\MI}{I}
\newcommand{\MIann}{I_{A,R}^{\mathrm{(ann)}}}
\newcommand{\T}{T}

\newcommand{\tscr}{t_{\mathrm{scr}}}

\newcommand{\pd}{p_d} % defining critical point for the disentangling transition

\begin{document}
\title{Quantum Coding Transitions in the Presence of Boundary Dissipation}
\author{Izabella Lovas}
\author{Utkarsh Agrawal}
\affiliation{Kavli Institute for Theoretical Physics, University of California, Santa Barbara, CA, 93106}
\author{Sagar Vijay}
\affiliation{Department of Physics, University of California, Santa Barbara, CA 93106}
\begin{abstract}
{We investigate phase transitions in the encoding of quantum information in a quantum many-body system due to the competing effects of unitary scrambling and boundary dissipation.  Specifically, we study the fate of quantum information in a one-dimensional qudit chain, subject to local unitary quantum circuit evolution in the presence of  depolarizating noise at the boundary.  If the qudit chain initially contains a finite amount of locally-accessible quantum information, unitary evolution in the presence of boundary dissipation allows this information to remain partially protected when the dissipation is sufficiently weak, and up to time-scales growing linearly in system size $L$.  In contrast, for strong enough dissipation, this information is completely lost to the dissipative environment.  We analytically investigate this ``quantum coding transition" by considering dynamics involving Haar-random, local unitary gates, and confirm our predictions in numerical simulations of Clifford quantum circuits.   We demonstrate that scrambling the quantum information in the qudit chain with a unitary circuit of depth $ \mathcal{O}(\log L)$ before the onset of dissipation can perfectly protect the information until late times. The nature of the coding transition changes when the dynamics extend for times much longer than $L$. We further show that at weak dissipation, it is possible to code at a finite rate, i.e. a fraction of the many-body Hilbert space of the qudit chain can be used to encode quantum information. }

\end{abstract}
\date{\today}
\maketitle

\section{Introduction}

% Quantum Coding Transition

The chaotic unitary evolution of an isolated quantum systems will spread initially localized quantum information over non-local degrees of freedom, a process known as quantum information scrambling~\cite{Hayden2007,Sekino2008,Nima_stanford,Hosur2016}. This delocalization of information aids in protecting quantum information against external interference from local noise, which is present in any real physical system. 
Studying the robustness of quantum information in the presence of both unitary scrambling and dissipation is important both to understand new dynamical regimes of quantum many-body dynamics, and from a practical standpoint, to design quantum codes and to appropriately interpret studies of quantum many-body evolution in near-term quantum simulators.
While dissipative dynamical phases of matter have been the subject of intense research for decades~\cite{Carr_book,Breuer_book,Zoller_review,Quantfluid_RevModPhys.85.29,Daley_review}, addressing the dynamics of quantum information in this context opens a new perspective. Similarly to how understanding the spreading of information has led to a deeper understanding of quantum chaos and thermalization~\cite{Sekino2008,Maldacena2016,Xu_2020,Enatanglement_RUC,nahum2018operator,OTOCS_RUC,Mezei_2017,Luitz2017}, studying quantum information in dissipative systems can shed light on the structure of (possibly new) dynamical regimes of quantum matter.%, and reveal new types of phases.

Besides its fundamental relevance for the dissipative dynamics of generic quantum systems, the fate of quantum information in the presence of unitary scrambling and destructive local noise or measurements has been explored in the context of quantum information theory, leading to the development of the theory of quantum error correcting codes~\cite{Shor_PhysRevA.52.R2493,Steane_PhysRevLett.77.793,Bennett_PhysRevA.54.3824,Knill_PhysRevA.55.900}. A key result in the theory of quantum error correction (QEC) is the threshold theorem, stating that for error rates below some threshold, one can reverse the effects of the errors by applying additional quantum gates~\cite{Knill_QEC,Aharonov_QEC,Shor_QEC}. In other words, it is possible to correct errors faster than they are created.

The threshold theorem is essential in designing fault-tolerant quantum computers. Applying additional gates, trying to preserve the code-space against the noise, allows one to perform logical operations for long times with high precision.  
Such an active error correction is feasible in artificial quantum systems with a ``digital" architecture, in which real-time measurements and unitary evolution can be executed over targeted degrees of freedom. 
However, in analog quantum simulators realized, e.g., with ultracold atoms, the options for active error correction are more restricted and costly due to the limited control over the dynamics. This provides a strong motivation for exploring whether the system's intrinsic dynamics alone can protect information, by  hiding it from destructive local noise. Despite this fundamental relevance, the conditions for obtaining such a robust, self-generated coding dynamics in a generic quantum system without any degree of external control, are still not fully explored.

Recently, the robustness of a self-generated code space against a special class  of  local perturbations has been investigated, taking the form of local projective measurements. These studies revealed a phase transition driven by the measurement rate, such that the code space can store an extensive amount of information, as long as the rate of measurements remains  below a finite threshold~\cite{Li_Chen_Fisher,Skinner_MIPT,Chan_MIPT,Gullans_Huse,Choi_MIPT}. However, this result cannot be generalized to more generic noise channels. For example, a quantum many-body system evolving in the presence of random erasures occurring in the bulk with finite rate destroys all quantum information in constant time ~\cite{Noh2020efficientclassical,Li_Sang_Hsieh_2023}, and active error-correction during the dynamics is required to protect the information beyond this time scale.  Understanding the conditions (if any) that  unitary evolution and local errors have to satisfy to guarantee the emergence of a robust, self-generated code space, without the need for an active error correction during the dynamics, is an open question of utmost relevance.

\begin{figure*}
    \centering
    $\begin{array}{cc}
         \includegraphics[width=0.45\linewidth]{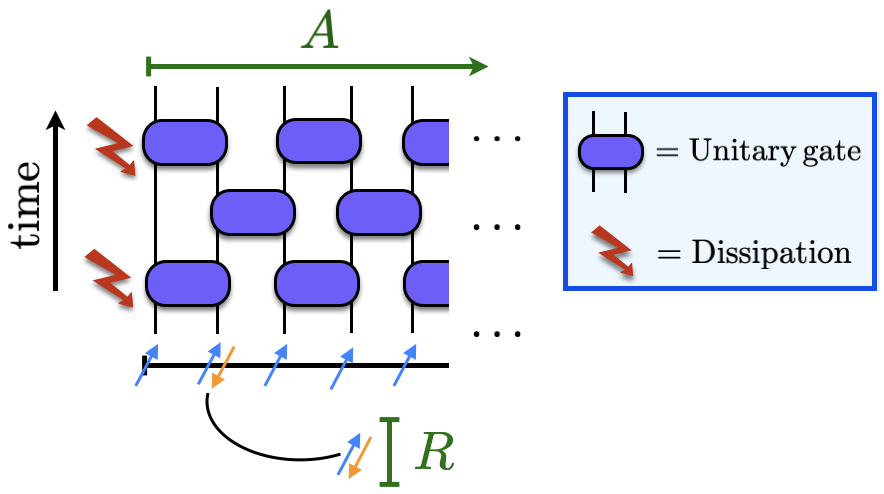}  &
         \includegraphics[width=0.4\linewidth]{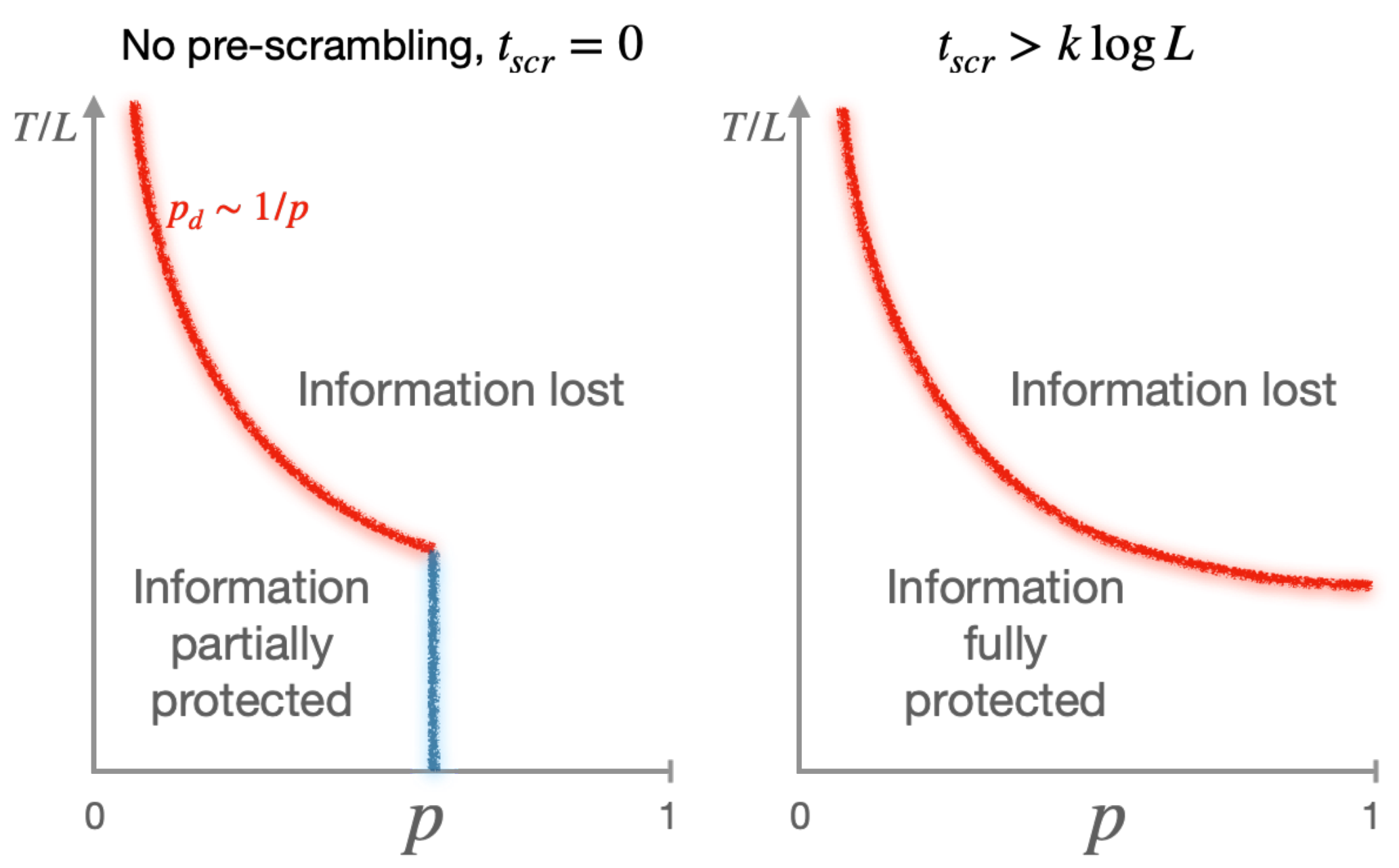} \\
         (a) & (b)
    \end{array}$ 
    \caption{(a) Quantum information is encoded in a qudit chain which subsequently evolves with a ``brickwork" array of Haar-random, two-site unitary gates and dissipation at the boundary. One timestep of these dynamics corresponds to two layers of unitary gates along with depolarizing noise at the boundary, as shown schematically in (a). A phase diagram for the {coding transition} is shown in (b). The blue critical line is the coding transition when the total number of timesteps $\T\lesssim L/p$, see Section~\ref{sec: coding transition}. This transition also corresponds to the de-pinning transition of an Ising domain wall in a statistical mechanical description of quantum information in these dynamics, as derived in the main text (Section~\ref{sec:I}). This transition occurs when $R$ is localized near the boundary and is not scrambled across the system. The red critical line is the coding transition as the system approches thermalization (see Section~\ref{sec:disentangling_transition}), across which the system becomes maximally entangled with the environment resulting in information loss.}
    
    \label{fig: summary_figure}
\end{figure*}

\subsection{Summary of Results} 
With these motivations, we take a step  towards understanding the dynamics of quantum information under generic scrambling and local noise, by exploring the fate of quantum information, subjected to the competing effects of boundary dissipation and unitary spreading in  a one-dimensional chaotic quantum system. For concreteness and simplicity, we focus on the setup sketched in Fig.~\ref{fig: summary_figure}a, which shows a single timestep of a random quantum circuit with a depolarization channel acting at the left boundary. We note that it is known both in classical coding theory~\cite{Shannon,Gallager_1962,Gallager1973,CodesDavid1996NearSL,richardson_2001_effcient} and in the quantum case~\cite{Brown_2013,Brown_2015,Gullans_2021_encoding}  that random unitary dynamics provides an optimal encoding of information. We entangle one external reference qubit $R$ near the boundary into a Bell pair, thereby encoding one qubit of quantum information initially localized near the dissipative boundary. We then ask what happens to this information as the system is subject to noisy dynamics, up to time scales $T$ scaling linearly with the system size $L$, such that $T/L$ is fixed. Importantly, by taking the thermodynamic limit $L\to\infty$ and the long time limit $T\to\infty$ simultaneously, with $T/L$ constant, we probe the system on time scales where it is expected to thermalize. 

Interestingly, we find that this quantum information can remain robust even at these long times, giving rise to a rich dynamical phase diagram as a function of dissipation strength $p$ and the ratio $T/L$, as displayed in Fig.~\ref{fig: summary_figure}b. The left panel shows the case where the noisy dynamics starts immediately after the encoding of the quantum information locally, near the leftmost boundary. We find a dissipation-induced quantum coding phase transition, separating a region where the coherent information remains partially  protected and gets delocalized within the system, and a phase where all of this information leaked to the environment. The nature of the coding transition, however, depends on the ratio $T/L$. For $T/L\lesssim 1$ the right boundary is effectively decoupled from the dynamics of information and we observe a continuous second-order phase transition (blue line). For even larger ratios $T/L$, the right boundary plays a crucial role and gives rise to a first order phase transition (red). We also demonstrate that adding a unitary ``pre-scrambling" step after the local encoding, before the onset of the dissipative dynamics, can efficiently increase the robustness of the encoded information. In particular, as shown in the right panel of Fig.~\ref{fig: summary_figure}b, a pre-scrambling time $t_{scr}$  scaling logarithmically with system size, $t_{scr}\sim\log L$, ensures that quantum information remains perfectly protected for small enough dissipation strengths $p$, up to time scales $T\sim L/p$.

We gain a detailed understanding of these different types of coding transitions, by mapping the dynamics of quantum information in a circuit with Haar-random unitary gates and boundary dissipation to the statistical mechanics of a two-dimensional lattice magnet.  This mapping, which has been extensively employed to understand unitary circuit quantum dynamics as well as dynamics with projective measurements (see Ref. \cite{fisher2022random, Potter2021} for a review), allows us to obtain analytical predictions, as well as instructive numerical results. While the entanglement measures of interest which diagnose the quantum coding transition require taking a formal replica limit of this lattice magnet (akin to a limit arising when considering ``quenched" disorder), we focus our attention on understanding this lattice magnet away from the replica limit (akin to studying an ``annealed" disorder-average).  Specifically, we focus on the ``annealed'' disorder average of the second R\'enyi mutual information between the output of the circuit $A$, and the reference qubit $R$. In this limit, the circuit with the boundary depolarization can be mapped to the statistical mechanis of an Ising magnet, in which a single Ising domain wall  experiences an attractive/repulsive potential at one boundary of the two-dimensional system, whose strength is tuned by the dissipation strength.  In this language, the coding transition at times $T/L\lesssim 1$ can be understood as a second order pinning/depinning transition of the Ising domain wall at the noisy boundary; we provide conjectures as to the true nature of this transition in the replica limit. At later times $T/L > 1/p$, the right boundary gives rise to a different, first order transition by ``absorbing" the Ising domain wall. Insights gained from this classical statistical picture are confirmed by large scale numerical simulations performed on Clifford quantum random circuits.

Finally, we show that the coding transition for $T/L > 1/p$ can also be understood as a transition arising from the monogamy of entanglement. In this case, as the system of $L$ qubits becomes entangled with a growing number of environmental degrees of freedom, scaling as $p T$, eventually it can no longer stay simultaneously entangled with the reference qubit, and all information leaks to the environment. We conclude with the interesting scenario of encoding an extensive amount of information in the system. Specifically, we show that a similar coding transition persists when we entangle an extensive number of reference qubits into Bell pairs with the qubits of the system. In particular, we identify two threshold values  for the dissipation strength $p$,  $p_{th,1}$ and $p_{th,2} $, separating three regions according to the behavior of the information density. The  information density  is perfectly protected in the system for $p<p_{th,1}$, while it starts to leak into the environment  above this threshold. A finite density of information still survives in the region $p_{th,1}<p<p_{th,2}$, until eventually reaching zero at the upper threshold $p_{th,2}$.

The rest of the paper is organized as follows. In Sec.~\ref{sec:I}, we introduce the mapping between the coherent quantum information in random circuits and the properties of an Ising domain wall experiencing a repulsive/attractive boundary on the left and an absorbing boundary on the right, by considering the ``annealed'' second R\'enyi mutual information between the circuit output and the encoded information. We derive the random walk model in Sec.~\ref{subsec:stat_mech}. We then show  in Sec.~\ref{subsec:encoding} that different phases on either side of the coding transition can be understood by inspecting the weighted trajectories of the Ising domain wall in this statistical mechanical model. 

We turn to the detailed discussion of the second order coding transition in the regime $ \T\lesssim L/p $, induced by the dissipative boundary alone without the interference of the clean boundary, in Sec. ~\ref{sec: coding transition}. We first rely on the random walk model to gain a qualitative understanding of the phase transition, and discuss the classical pinning/depinning transition of the Ising domain wall in Sec.~\ref{subsec:annealed_MI}. Building on these insights, we verify the presence of the quantum coding transition and  study its properties numerically in Sec~\ref{subsec: CliffordMI}, by performing large scale numerical simulations on Clifford quantum circuits, before discussing the nature of this transition in more detail in Sec.~\ref{sec:phase_transition}. To end the section, in Sec.~\ref{subsec:logscrambling} we comment on increasing the robustness of the encoded information by applying a unitary pre-scambling before the onset of dissipative dynamics. We show that a pre-scrambling time $ \tscr$ scaling logarithmically with system size provides perfect protection for the coherent information for weak enough dissipation $p$, up to time scales $T/L\sim O(1)$.

We turn to the first order coding transition, induced by the interplay of the dissipative left boundary and the clean right boundary at times $ \T\gtrsim L/p $, in Sec.~\ref{sec:disentangling_transition}. First, we discuss that this phase transition can be understood in the statistical mechanical framework as the absorption of the entanglement domain wall by the right boundary and is driven by the monogamy of entanglement as the system becomes entangled with a growing number of environmental qubits.	We present and analyze the numerical results obtained from Clifford circuit simulations in Sec.~\ref{subsec:disentangling_numerics}, and find  good agreement with the predictions of the statistical mechanics of the Ising lattice magnet. We argue that this coding transition is of first order, and discuss its scaling properties in Sec.~\ref{subsec:disent_discussion}. Finally, Sec~\ref{sec:finite_code_rate} serves as an outlook to the case of encoding an extensive amount of information into the system. Here we consider entangling a finite density of reference qubits with the system, and find a monogamy induced coding transition at late times $ \T\gtrsim L/p $, similar to the one observed for a single bit of quantum information. Here we find three phases, with the information perfectly protected for $p<p_{th,1}$, a finite density of information surviving for $p_{th,1}<p<p_{th,2}$, and the density reaching zero above $p_{th,2}$. We conclude by summarizing our results, and discussing open questions in Sec.~\ref{sec:discuss}.

\tableofcontents

\section{Dissipation in Quantum Circuit Evolution}\label{sec:I}

\subsection{Statistical Mechanics of Random Unitary Evolution and Dissipation}\label{subsec:stat_mech}
Past studies of {random} local unitary evolution \cite{fisher2022random,Potter2021}, evolution with projective measurements~\cite{Li_Chen_Fisher,Skinner_MIPT,Chan_MIPT} and with dissipation~\cite{Li_Sang_Hsieh_2023,Noh2020efficientclassical,Jian2021_Noisy,Yaodong_Fisher_Decoding,Lucas_Noisy_2020,Weinstein_2022} have uncovered a wealth of universal structures governing the dynamics of information-theoretic quantities such as the R\'{e}nyi entanglement entropy.  Averaging over an ensemble of unitary gates in this setting gives rise to an emergent classical statistical mechanics of quantum entanglement, which must be understood in an appropriate ``replica limit" in order to recover the behavior of the information-theoretic quantities of interest.  A qualitatively-accurate understanding of the behavior of quantum entanglement in chaotic unitary dynamics, and in dynamics with projective measurements can still be obtained even without taking the replica limit \cite{fan2021self,li2021statistical,nahum2018operator,Bao2020}, though these approaches often fail to capture quantitative, universal properties characterizing distinct regimes of quantum many-body evolution (e.g. of the volume-law-entangled phase of infrequently monitored quantum many-body evolution \cite{li2023entanglement}) or of critical points (e.g. separating different phases of monitored quantum dynamics).    

Here, we consider the evolution of qudits under random, local unitary gates and boundary dissipation.  Averaging over the ensemble of unitary gates, in the calculation of the evolving \emph{purity} of subsystem, leads to an emergent statistical mechanics of an Ising magnet.  We present the various ingredients that the unitary evolution and dissipation correspond to in this setting, before using these ingredients extensively in subsequent sections to understand the stability of encoded quantum information under this evolution.

\begin{figure}
$\begin{array}{c}
              \includegraphics[width=0.38\textwidth]{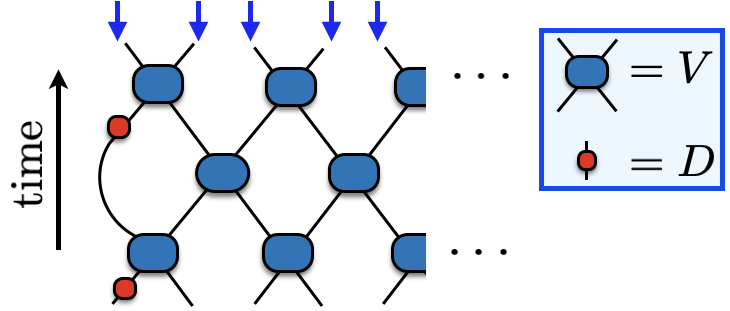} \\
              \includegraphics[width=0.27\textwidth]{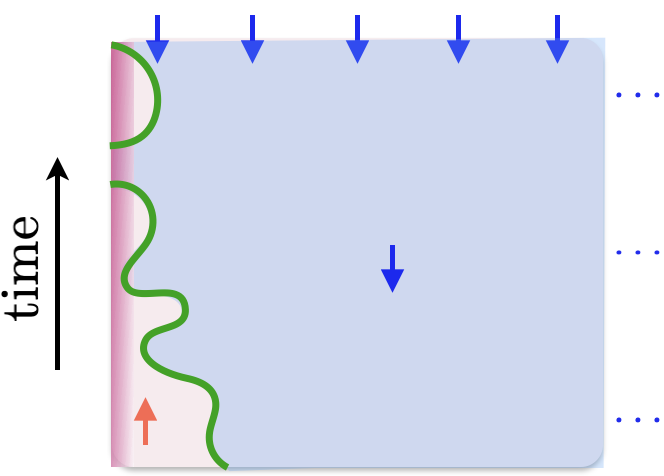}\\
              %\text{(a)} & \text{(b)} &\text{(c)}
\end{array}$
             \caption{\textit{Top.} Performing a Haar-average over the unitary gates in the calculation of the purity of the evolving state gives rise to an Ising magnet, whose partition function may be written as the product of transfer matrices, given in Eq. (\ref{eq:Haar_transf_1}), (\ref{eq:Haar_transf_2}) and  (\ref{eq:D}).  \textit{Bottom.} A coarse-grained description of this Ising magnet involves a single Ising domain wall (green)in the presence of a boundary magnetic field (shaded red). The boundary conditions at the bottom of the Ising magnet, which are fixed by the initial state of the quantum system, are not shown. }
         \label{fig:Circuit}
\end{figure}

We focus our attention on a one-dimensional chain of qudits, with Hilbert space dimension $q$ at each lattice site.  The dissipation acts on the boundary qudit, and is described by the depolarizing channel $\Phi$ acting on the density matrix $\rho$ of this qudit as 
\begin{align}\label{eq:depolarize}
\Phi(\rho) = (1-p)\,\rho + p\cdot \frac{\mathds{1}_{q\times q}}{q}
\end{align}
with $p\in[0,1]$ parametrizing the ``strength" of the dissipation.  For future convenience, we choose to rewrite the depolarizing channel as an \emph{operator} $\hat{\Phi}$ which acts within a Hilbert space of dimension $q^{2}$.  The operator $\hat{\Phi}$ takes the form
\begin{align}\label{eq:dissip_operator}
\hat{\Phi} = \sum_{i,j = 1}^{q}\left[(1-p)\ket{i,j}\bra{i,j} + \frac{p}{q}\ket{i,i}\bra{j,j}\right]
\end{align}
where $\ket{i}$ for $i\in\{1,\ldots,q\}$ denotes an orthonormal basis of states of a single qudit\footnote{The qudit density matrix $\rho \equiv\sum_{i,j}\rho_{ij}\ket{i}\bra{j}$ is a \emph{state} $\ket{\rho} \equiv \sum_{i,j}\rho_{ij}\ket{i,j}$ in the doubled Hilbert space on which the operator $\hat{\Phi}$ acts as $\hat{\Phi}\ket{\rho} = (1-p)\ket{\rho} + (p/q)\sum_{i}\ket{i,i}$. }.

Apart from the dissipation, the remaining qudits will be chosen to evolve according to two-site unitary gates, chosen from the uniform (Haar) measure for the unitary group U$(q^{2})$.  Given such a two-qudit unitary gate $U$, we note that the average over the Haar measure of $U\otimes U^{*}\otimes U\otimes U^{*}$ -- a quantity which will naturally appear in subsequent sections -- is given by
\begin{align}
V\equiv &\langle U\otimes U^{*}\otimes U\otimes U^{*}\rangle\nonumber\\
&= \sum_{\sigma,\tau\in\{\uparrow,\downarrow\}}\mathrm{wg}_{2}(\sigma\tau)\ket{\tau,\tau}\bra{\sigma,\sigma}\label{eq:Haar_unitary}
\end{align}
where $\langle\cdots\rangle$ denotes the Haar average, the Weingarten function is given as 
$\mathrm{wg}_{2}(+) = \frac{q^{2}}{q^{4}-1}$ and $\mathrm{wg}_{2}(-) = \frac{-1}{q^{4}-1}$, 
and the states $\ket{\uparrow}$ and $\ket{\downarrow}$ are defined as
 $\ket{\uparrow} \equiv \sum_{i,j = 1}^{q}\ket{i,i,j,j}$ and
$\ket{\downarrow} \equiv \sum_{i,j = 1}^{q}\ket{i,j,j,i}$ so that 
\begin{align}\label{eq:overlap}
\braket{\sigma |\tau} = (q^{2}-q)\delta_{\sigma,\tau} + q. 
\end{align}
From these expressions, it is clear that 
\begin{align}
&V\ket{\uparrow \uparrow} = \ket{\uparrow\uparrow}\hspace{.7in} V\ket{\downarrow \downarrow} = \ket{\downarrow\downarrow}\label{eq:Haar_transf_1}\\
&V\ket{\uparrow\downarrow} = V\ket{\downarrow\uparrow} = \frac{q}{q^{2}+1}\left[\ket{\downarrow\downarrow} + \ket{\uparrow\uparrow}\right]\label{eq:Haar_transf_2}
\end{align}
From Eq. (\ref{eq:dissip_operator}), the operator $D\equiv \hat{\Phi}\otimes\hat{\Phi}$ acts on these states as 
\begin{align}\label{eq:D}
D\ket{\uparrow} = \ket{\uparrow} \hspace{.3in} D\ket{\downarrow} = (1-p)^{2}\ket{\downarrow} + \frac{p(2- p)}{q}\ket{\uparrow} 
\end{align}

\subsection{Boundary Dissipation and the Encoding of Quantum Information}\label{subsec:encoding}
We now consider a qudit chain consisting of $L$ qudits, into which quantum information has been encoded.   We may imagine that this quantum information is represented by physical reference qudits which are maximally-entangled with the one-dimensional system.  This system subsequently evolves according to a unitary circuit composed of Haar-random unitary gates in a ``brickwork" array, together with dissipation which acts near the boundary.  We first focus on the case where only a single qudit is encoded in the one-dimensional system, and with dissipation acting periodically in time on the boundary qudit, as shown schematically in Fig. \ref{fig:Circuit}a.  A single timestep of this evolution corresponds to the application of two layers of two-site unitary gates, followed by the depolarizing channel (\ref{eq:depolarize}) on the boundary qudit.

To diagnose whether this qudit of encoded information can remain in the system, even as the boundary dissipation continues to act, we study the behavior of the bipartite mutual information between the reference qudit ($R$), and the system ($A$) at a time $t$; this mutual information is defined as
\begin{align}\label{eq:I}
I_{A,R}(t) = S_{A}(t) + S_{R}(t) - S_{A\cup R}(t)
\end{align}
where $S_{A} \equiv -\Tr\left[\rho_{A}(t)\log_{q}\,\rho_{A}(t)\right]$ is the von Neumann entanglement entropy of subsystem $A$ at a time $t$.  We note that $\MI_{A,R}(t)$  is related to the coherent information present in the system. If $\MI_{A,R}=2$ the entangled qudit can be perfectly recovered by applying a recovery operation to the system \textit{alone} whereas for $\MI_{A,R}=0$ the information has leaked to the environment, that is, $\MI_{E,R}=2$~\cite{Schumacher1996,Schumacher2001}.

 The mutual information (\ref{eq:I}) averaged over realizations of the random unitary evolution, thus diagnoses whether quantum information remains in the system, even in the presence of  boundary dissipation.  Instead of considering the Haar-average of the mutual information, we turn our attention on the ``annealed" average of the second R\'{e}nyi mutual information between $A$ and $R$, defined as 
\begin{align}\label{eq:annealed_I}
I_{A,R}^{(\mathrm{ann})}(t) \equiv \log_{q}\langle q^{\,I_{A,R}^{(2)}(t)}\rangle
\end{align}
where $I_{A,R}^{(2)}(t) = S_{A}^{(2)}(t) + S_{R}^{(2)}(t) - S_{A\cup R}^{(2)}(t)$, with the second R\'{e}nyi entropy defined as $S_{A}^{(2)} \equiv - \log_{q}\mathrm{Tr}\rho_{A}(t)^{2}$, and $\langle\cdots\rangle$ denotes the Haar average over the unitary gates in the circuit.  The behavior of the annealed mutual information (\ref{eq:annealed_I}) can provide a qualitative understanding of the quantity of interest (\ref{eq:I}), as discussed at the beginning of this section, though quantitative details may differ, as we will later clarify.  

We proceed to calculate the annealed mutual information (\ref{eq:annealed_I}). We initialize the qudits in a product state, except for the qudit at a site $x_{0}$ away from the boundary which is maximally entangled with the reference qudit.  As the system evolves in the presence of unitary gates and dissipation, it is evident that the purity of the reference qudit remains unchanged, $\Tr\rho_{R}(t)^{2} = q^{-1}$ for all times $t$.  Furthermore,  calculation of $\langle\Tr\rho_{A}(t)^{2}\rangle$ and $\langle\Tr\rho_{A\cup R}(t)^{2}\rangle$ involves performing a Haar average of four copies of the quantum circuit.  Following the discussion in the previous section, it is thus clear that these Haar-averaged purities may be written as partition functions for an Ising magnet of finite extent in the vertical direction -- corresponding to the time direction in the quantum circuit -- and with horizontal extent fixed by the number of qudits in the system.  The Ising spins live on the links of a square lattice, and are acted upon by the transfer matrices matrices $V$ and $D$, as given in Eq. (\ref{eq:Haar_transf_1}), (\ref{eq:Haar_transf_2}) and  (\ref{eq:D}), depending on whether a Haar-random unitary gate or dissipation is applied at a particular point in spacetime in the quantum circuit, respectively.  The full transfer matrix is shown schematically in Fig. \ref{fig:Circuit}b.

The boundary conditions for the Ising partition sum, at the ($i$) bottom and ($ii$) top boundaries are determined by ($i$) the initial state of the qudit chain along with the location of the reference qudit, and ($ii$) the subsystem over which the purity is being calculated, respectively. First, fixing Ising spins at the top boundary to be in the $\downarrow$ state corresponds to keeping the corresponding qudit within the region for which the purity is being calculated.  As a result, the spins at the top boundary are all fixed in the $\downarrow$ state for both 
the calculation of $\langle \Tr\,\rho_{A}(t)^{2}\rangle$ and $\langle \Tr\,\rho_{A\cup R}(t)^{2}\rangle$, as shown in Fig. \ref{fig:Circuit}b. These two purities thus only differ in their bottom boundary conditions.  Here, the boundary spins are allowed to freely fluctuate, with the exception of the spin corresponding to the qudit at a distance $x$ away from the boundary; the state of this Ising spin determines whether the reference qudit is included in the subsystem whose purity is being computed.  More precisely, this spin is fixed in the $\uparrow$ or $\downarrow$ state in the calculation of the quantities, $\langle \Tr\,\rho_{A}(t)^{2}\rangle$ and $\langle \Tr\,\rho_{A\cup R}(t)^{2}\rangle$, respectively.

It is convenient to evaluate these partition functions by contracting the transfer matrix from the top boundary condition, i.e. ``backwards" in time with respect to the arrow of time in the quantum circuit. Let $Z(t)$ denote the partition sum obtained by evolving the all-down state of the Ising spins for $t$ timesteps by repeatedly applying the row transfer matrix corresponding to a single timestep of the dynamics.  
The partition sum $Z(t)$ describes a single, directed Ising domain wall, which can only be created/annihilated at the boundary of the system.  This can be seen as follows.  First, starting with the all-down state, the dissipation (\ref{eq:D}) can flip the boundary Ising spin from $\ket{\downarrow}$ to $\ket{\uparrow}$, thus creating an Ising domain wall near the boundary.  The effect of the Haar-random unitary gates (\ref{eq:Haar_transf_1}), (\ref{eq:Haar_transf_2}) in the bulk of the quantum circuit is to simply move the domain wall.  Notably, Eq. (\ref{eq:Haar_transf_1}) implies that the Haar-random gates cannot create or annihilate Ising domain walls in the bulk of the system, though gates acting near the boundary can annihilate the Ising domain wall.  Once the state of the boundary spin is $\ket{\uparrow}$, the dissipation cannot alter this state since $D\ket{\uparrow} = \ket{\uparrow}$; this is simply a consequence of the fact that the depolarizing channel (\ref{eq:depolarize}) leaves the maximally-mixed density matrix $\rho = \mathds{1}_{q\times q}/q$ unchanged.  

The partition sum $Z(t)$ is thus performed over histories of the entanglement domain wall trajectories, which can propagate in the bulk of the system, or be created/annihilated at the boundary.  Formally, we write 
\begin{align}
Z(t) = \sum_{x\ge 0}z(x,t)
\end{align}
where $z(x,t)$ is a restricted sum over trajectories of the entanglement domain wall where the domain wall ends up between sites $x-1$ and $x$ at time $t$. In this convention, $z(0,t)$ corresponds to trajectories where the entanglement domain wall no longer exists at time $t$, as it has been annihilated at the left interface.

We may now write the Haar-averaged purities as 
\begin{align}
\langle \Tr\,\rho_{A}(t)^{2}\rangle = q^{2}\sum_{y>x_{0}}z(y,t) + q\sum_{y\le x_{0}}z(y,t)\\
\langle \Tr\,\rho_{A\cup R}(t)^{2}\rangle = q^{2}\sum_{y\le x_{0}}z(y,t) + q\sum_{y>x_{0}}z(y,t)
\end{align}
This is due to the fact that $\langle \Tr\,\rho_{A\cup R}(t)^{2}\rangle$  involves a sum over trajectories of the entanglement domain wall, with an additional weight $q^{2}$ given to trajectories which end at a position $y > x_{0}$ and a weight $q$ given to trajectories ending at $y \le x_{0}$, where $x_{0}$ is the location of the entangled reference qudit. The opposite weighting scheme is true for $\langle\Tr\,\rho_{A}(t)^{2}\rangle$. These additional weights arise due to the fact that depending on the final position of the entanglement domain wall, the boundary spin at $x$ is contracted with the state $\ket{\uparrow}$ or $\ket{\downarrow}$.  These overlaps are given in Eq. (\ref{eq:overlap}).  With these expressions, it is straightforward to see that 
\begin{align}
I_{A,R}^{(\mathrm{ann})}(t) = \log_{q}\left[\frac{q^{2} - q(q-1)P(x_{0},t)}{1 + (q-1)P(x_{0},t)}\right] \label{eq:annealed_MI}
\end{align} 
where
\begin{align}
P(x_{0},t) \equiv \frac{1}{Z(t)}\sum_{y\ge x_{0}}z(y,t)
\end{align}
is the probability that the domain wall ends at a position $y\ge x_{0}$ at time $t$.  

\section{Quantum Coding Transition} \label{sec: coding transition}
In this section, we study the behavior of the encoding of quantum information in the system, after evolving the system by the quantum circuit for $T$ timesteps, for a fixed dissipation strength $p$.  The number of timesteps of the evolution $T$ can be large so that $T/L \sim O(1)$ but is taken to be small enough throughout the entirety of this section, so that the left and right ends of the one-dimensional qudit chain are causally disconnected.  As $p$ is increased from zero, we will find an ``quantum coding" transition, where information initially encoded in the system is lost to the environment above a threshold $p = p_{c}$.  

\subsection{Annealed Mutual Information, and the Pinning of an Ising Domain Wall}\label{subsec:annealed_MI}
First, we investigate the behavior of  $I_{A,R}^{(\mathrm{ann})}$ as the dissipation strength $p$ is tuned, by studying the Ising lattice magnet that emerges after performing a Haar-average over the unitary gates in the quantum circuit.  

As discussed in Sec. \ref{subsec:encoding}, the partition sum $Z(T)$ describes a single Ising domain wall which can propagate through the bulk of the two-dimensional system, and be created/annihilated at the left boundary of the system.  Tuning the dissipation strength, which alters the Ising symmetry-breaking field applied at the boundary, modulates an effective ``pinning potential" for the Ising domain wall.  This can be clearly seen in the limiting cases when $p = 0$ or $1$.  In the former case, the dissipation is completely absent, and Eq. (\ref{eq:Haar_transf_1}) implies that the all-down state is left invariant by the transfer matrix for the Haar-averaged circuit.  Thus, in this limit, there is no Ising domain wall.  In contrast, when $p=1$, the boundary spin is fixed in the $\ket{\uparrow}$ state, and the domain wall is effectively repelled from the left boundary.  

Increasing the dissipation strength can then drive a pinning/de-pinning phase transition for the entanglement domain wall.  Similar phase transitions due to the presence of a boundary magnetic field in an Ising magnet have been studied in the literature (see, e.g. Ref. \cite{abraham1980solvable,abraham1981binding,chalker1981pinning}).  Equivalently, the temporally-directed nature of the Ising domain wall also suggests these paths may be thought of as the imaginary-time trajectories of a single quantum-mechanical particle on the half-line, which experiences a potential near the boundary, which is tuned by the dissipation strength. $Z(T)$ is thus an amplitude for this particle to propagate under imaginary time-evolution by this Hamiltonian. In this setting, the particle can undergo a localization transition when the potential is \emph{sufficiently} attractive \cite{abraham1981binding}.  This result is to be contrasted with the well-studied problem of a particle on the full line, with a delta-function potential near the origin, which always forms a bound-state in the potential well as long as the potential is attractive.

The annealed mutual information precisely measures the localization of the Ising domain wall, as is evident from Eq. (\ref{eq:annealed_MI}).  Deep within a localized phase, where the transverse wandering of the domain wall is governed by a length-scale $\ell_{\perp}$, the probability $P(x_{0},T) \sim e^{-x_{0}/\ell_{\perp}}$ ($\ell_{\perp}\ll x_{0})$, so that $I^{(\mathrm{ann})}_{A,R}$ is a constant, deviating from its maximal value of $2$ by a constant correction which changes within the localized phase.  In contrast, in the delocalized phase, the probability $P(x_{0},T) \overset{T\rightarrow\infty}{=} 1$, where the limit is taken, keeping the ratio $T/L=\mathrm{const.}$ fixed. 

\begin{figure}[t]
\includegraphics[width=.4\textwidth]{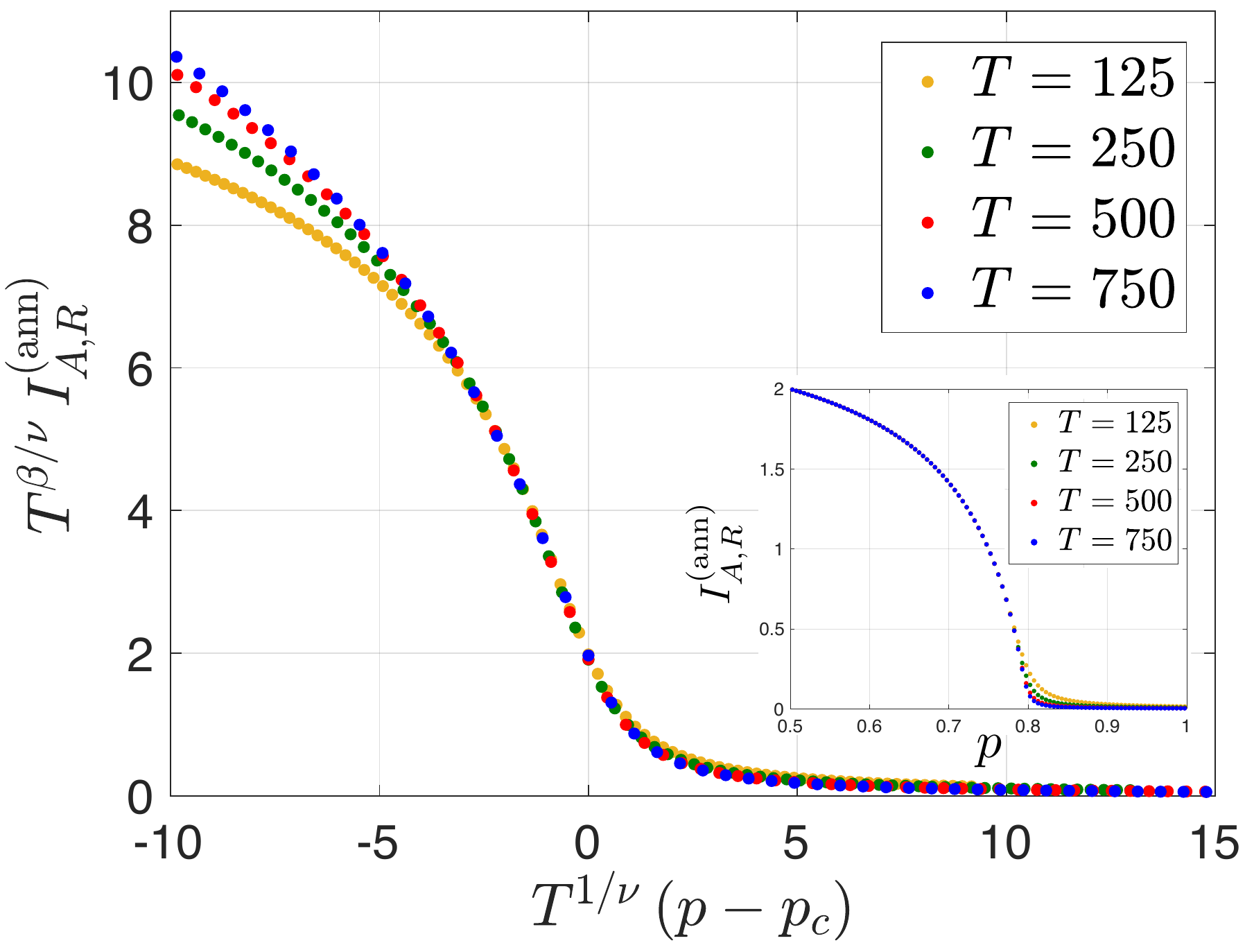}
             \caption{Scaling collapse of the annealed mutual information, consistent with the scaling form in Eq. (\ref{eq:annealed_scaling_form}).  The inset shows the behavior of the annealed mutual information as a function of dissipation strength $p$, indicating the presence of an coding transition.  The exponents $\beta = 1/2$, $\nu = 2$ are determined from properties of the pinning transition of the Ising domain wall.    The system size is taken to be large enough that the left and right ends of the qudit chain are causally disconnected.}
         \label{fig:Scaling_Collapse_Semi_Inf}
\end{figure}

Properties of this coding transition, as seen by annealed-averaged observables, such as the annealed mutual information, may be obtained by studying the lattice partition function for the Ising domain wall, which we present in Appendix \ref{app:lattice_partition}, due to the technical nature of the calculations involved.  From this study, we find that
\begin{enumerate}
    \item The phase transition occurs at a probability $p_{c}$ which varies as a function of the on-site Hilbert space dimension $q$.  The behavior of $p_{c}$ as $q$ is tuned may be determined by studying the lattice partition function.  In the limit $q\rightarrow\infty$, the coding transition is absent.  Specifically, we find that
    \begin{align}
        p_{c} = 1 - O(q^{-2})
    \end{align}
    so that information is always preserved in the system in the limit that the on-site Hilbert space dimension is strictly infinite.

    \item Near the phase transition, the annealed mutual information takes the universal scaling form 
    \begin{align}\label{eq:annealed_scaling_form}
        I^{(\mathrm{ann})}_{A,R}(T) = T^{-\beta/\nu}F(T^{1/\nu}(p-p_{c}))
    \end{align}
    where $\beta = 1/2$ and $\nu = 2$.  The function $F(x)\sim x^{\beta}$ as $x\rightarrow -\infty$. This relation is obtained by determining that in the thermodynamic limit, the annealed mutual information should vanish on approaching the transition as $I^{(\mathrm{ann})}_{A,R} \sim \ell_{\perp}^{-1}$, where $\ell_{\perp}$ is the distance of a transverse excursion of the Ising domain wall in the pinned phase.  This length scale is shown to diverge as $\ell_{\perp} \overset{p\rightarrow p_{c}^{-}}{\sim} (p_{c} - p)^{-\beta}$ upon approaching the phase transition.
\end{enumerate}
The above scaling form for the annealed mutual information is in good quantitative agreement with numerical studies, which we perform by directly studying the transfer matrix for the Ising magnet.  A numerically-obtained scaling collapse for the annealed mutual information is shown in Fig. \ref{fig:Scaling_Collapse_Semi_Inf}, which is consistent with Eq. (\ref{eq:annealed_scaling_form}).  

We expect that the qualitative behaviors presented here hold for the ``quenched-averaged" quantities of interest, such as the averaged von Neumann mutual information $\langle I_{A,R}(t)\rangle$, which truly diagnose the loss of quantum information from the system, as the dynamics proceed.  The true nature of the phase transition, however, will be different, as we discuss in Sec. \ref{sec:phase_transition}.

\subsection{Numerical Study}\label{subsec: CliffordMI}
Having obtained a qualitative understanding of the coding transition by considering the ``annealed" Haar average of the R\'enyi mutual information, we now demonstrate the presence of this transition in numerical studies of quantum circuit evolution in a qubit chain ($q=2$ on-site Hilbert space dimension). Here, the unitary time evolution of the bulk is governed by Clifford random unitary gates, arranged in a brickwork structure. This setup allows us to simulate the dynamics of large systems for sufficiently long times to study the phase transition introduced above, by relying on the stabilizer formalism. The boundary dissipation is realized as a random erasure channel, acting on  the leftmost qubit with probability $p$ in each time step, by deleting the information stored in the qubit. In the stabilizer formalism, this boundary erasure channel is implemented by deleting all stabilizers acting non-trivially (as a non-identity operator) on the leftmost qubit. 

We note that besides the protocol described above, we also considered other forms of boundary dissipation and Clifford scrambling, all giving rise to similar results for the behavior of the mutual information. Specifically, we implemented an alternative dissipation channel, by applying a CNOT gate entangling the boundary qubit with an environmental ancilla qubit that was subsequently traced out from the density matrix. Moreover, we considered protocols with sparse bulk scrambling, where each unitary gate in the brickwork structure is a random Clifford unitary with probability $p_U<1$, but the trivial identity operator with probability $1-p_U$.  This scenario allowed us to tune the efficiency of the scrambling through the parameter $p_U$, while keeping the boundary noise fixed, leading to a phase transition similar to the one discussed in the main text. We discuss these alternative protocols in more detail, and present supplementary numerical results in Appendix~\ref{app:protocols}.

The Bell pair is encoded in the initial state at the leftmost site, by entangling the boundary qubit with a reference qubit, while the remaining qubits are initialized in a random product state. We run the dissipative dynamics for time $T$, with system size $L$ chosen to keep  $T/L<1$ fixed, such that the right  boundary of the system is not causally connected to the Bell pair. This setting allows us to detect the coding transition induced by a single boundary, by increasing the evolution time $T$. Importantly, due to the fixed ratio $T/L$, the long time limit $T\to\infty$ and the thermodynamics limit $L\to\infty$ are performed simultaneously, therefore, we are probing the mutual information on time scales where the system is expected to become thermalized.

\begin{figure}
    \centering
    \includegraphics[width=.9\linewidth]{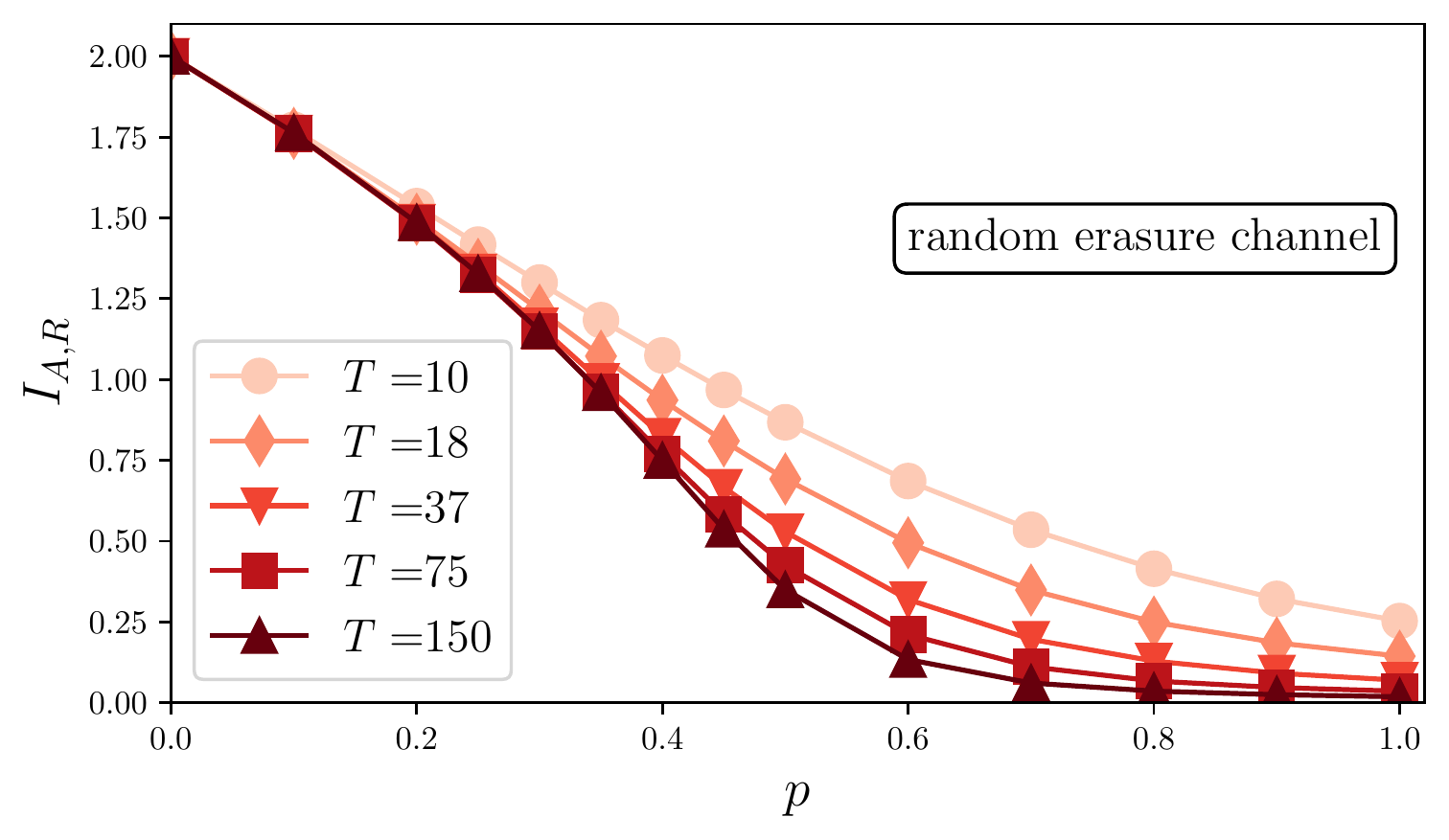} \\
    \caption{Coding transition induced by a single boundary. The mutual information between the reference qubit and the output of the circuit shown as a function of dissipation strength $p$, for $T / L<1$ fixed, with boundary dissipation realized as a random erasure channel. The scaling with circuit depths $T$ points to a phase transition between a phase with partially protected information, and a phase with all information lost.}
    \label{fig:I_smallT}
\end{figure}

\begin{figure}
    \includegraphics[width=.9\linewidth]{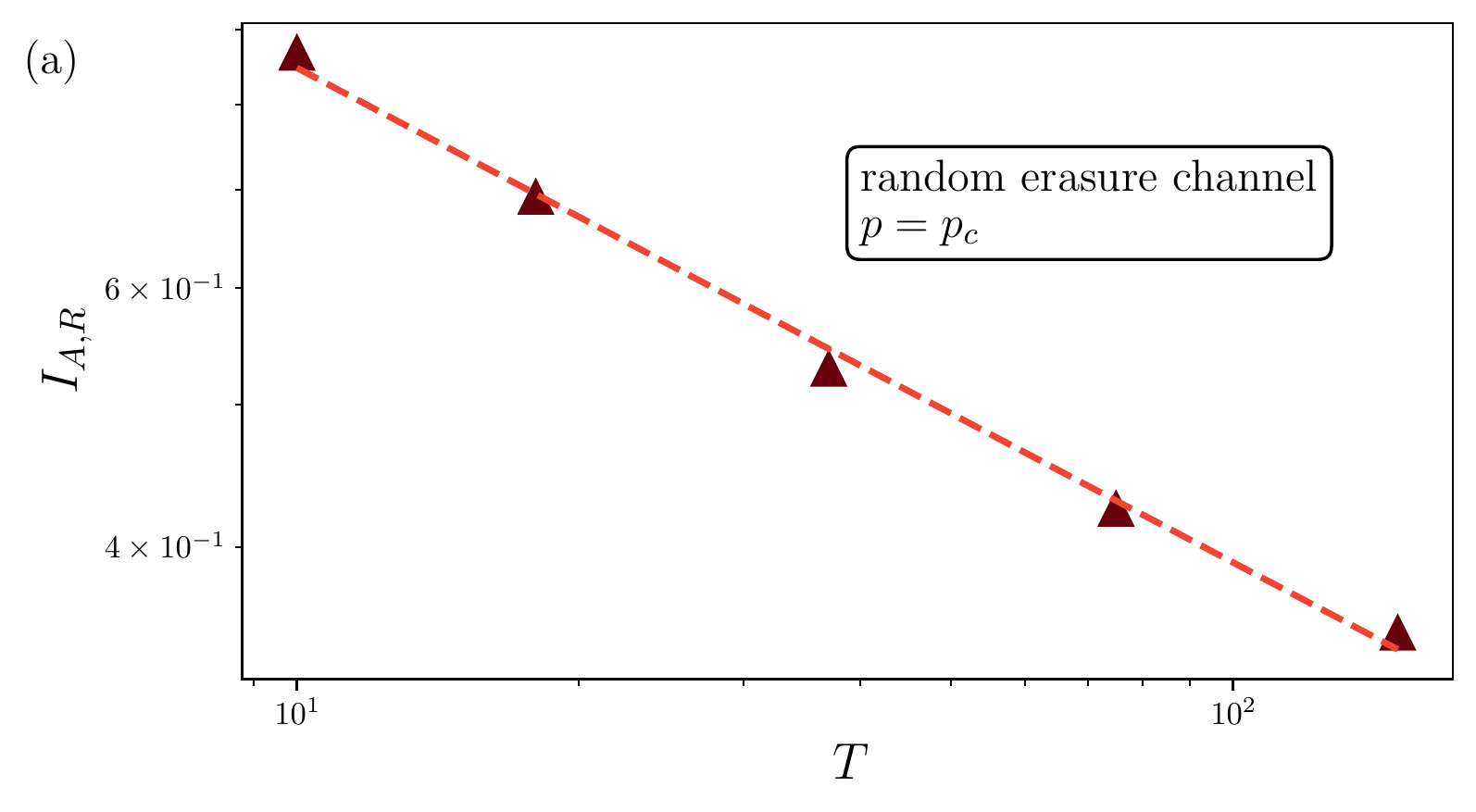} \\
    \includegraphics[width=.9\linewidth]{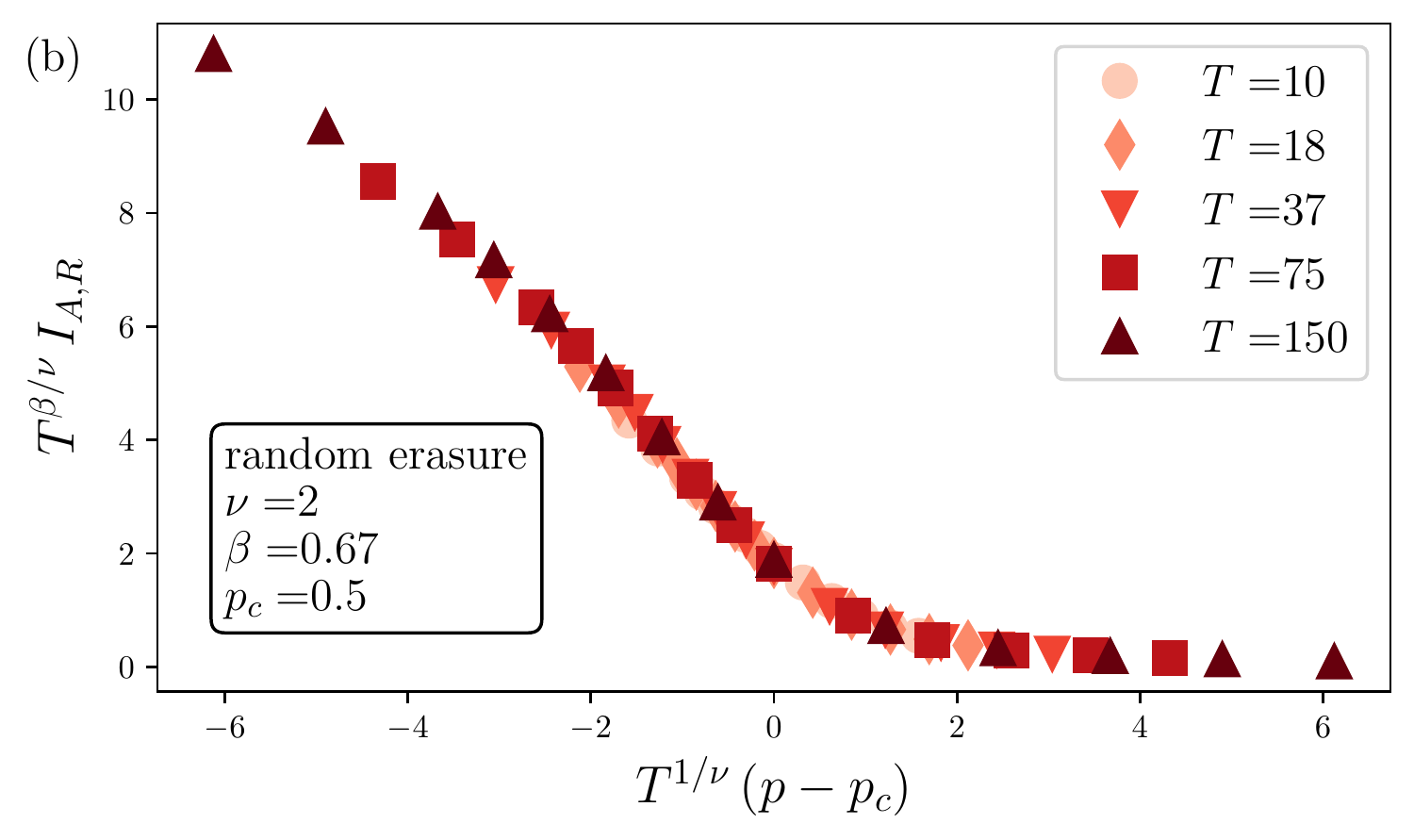}
    \caption{Critical properties of the coding transition for a single boundary. (a) Critical power law scaling of the mutual information with respect to circuit depth $T$ at the estimated transition point, $p_c=0.5$. The scaling relation $I_{A,R}\sim T^{-\beta/\nu}$ is used to extract $\beta/\nu=0.34$ (dashed line). (b) Full scaling collapse of rescaled mutual information $T^{\beta/\nu}I_{A,R}$ as a function of $T^{1/\nu}\left(p-p_c\right)$, using $\nu=2$.}
    \label{fig:Ipc}
\end{figure}

The mutual information $I_{A,R}$ between the output of the dissipative quantum circuit $A$ and the reference qubit $R$  is shown in Fig.~\ref{fig:I_smallT}, for different dissipation strengths $p$ and circuit depths $T$. These results are consistent with a coding transition tuned by the dissipation strength $p$, between a phase where the system retains part of the encoded information, and a strongly dissipative phase with all information lost. We note that determining the critical exponents and critical point of this transition from finite time data is numerically challenging. Nevertheless, we attempt to estimate these parameters by noting that the mutual information obeys the finite size scaling $I_{A,R}\sim T^{-\beta/\nu}$ at the critical dissipation strength $p_c$, while it saturates to a finite value as $T\rightarrow\infty$ for $p<p_c$. Relying on this observation, we identify $p_c$ with the smallest $p$ where the numerical data are consistent with $I_{A,R}$ approaching zero algebraically as $T\rightarrow\infty$, yielding the estimate $p_c\approx 0.5$. We then use the critical scaling $\left.I_{A,R}\right|_{p=p_c}\sim T^{-\beta/\nu}$ to fit the ratio $\beta/\nu$, see Fig.~\ref{fig:Ipc}a. Finally, we fit estimate $\nu$ by requiring a good scaling collapse for the full set of data from Fig.~\ref{fig:I_smallT}. We obtain the critical parameters $p_c=0.5$, $\beta/\nu=0.34$ and $\nu=2$, yielding the scaling collapse shown in Fig.~\ref{fig:Ipc}b. We note, however, that due to the large number of fitting parameters, the critical exponents extracted this way carry a considerable uncertainty. We leave the more thorough investigation of critical properties for future work.

\subsection{The Replica Limit and the Nature of the Phase Transition}\label{sec:phase_transition}

The behavior of quenched-averaged quantities, e.g. the Haar-averaged R\'{e}nyi mutual information $\langle I_{A,R}^{(2)}(t)\rangle$, close to the coding phase transition are quantitatively distinct from the annealed-averaged mutual information studied in Sec. \ref{subsec:annealed_MI}.   This is suggested by the numerical studies in the previous section, which present strong evidence that the coding phase transition is in a different universality class from a de-pinning phase transition for a single Ising domain wall. Here, we will provide some conjectures on the nature of this phase transition, based on analytic arguments.

We will focus our attention on the averaged second R\'{e}nyi mutual information $\langle I_{A,R}^{(2)}(t)\rangle$ whose behavior may be obtained via a ``replica trick"; the second R\'{e}nyi entropy may be obtained in the limit $S_{A}^{(2)}(t) = \displaystyle\lim_{k\rightarrow 0}\left(1-\left[\Tr\rho_{A}(t)^{2}\right]^{k}\right)/k$, so that the calculation of the Haar-averaged mutual information reduces to evaluating quantities such as $\langle\left[\Tr\rho_{A}(t)^{2}\right]^{k}\rangle$ in a replica limit $k\rightarrow 0$.  After the Haar average, these quantities may be regarded as partition functions for lattice magnets with ``spins" taking values in the permutation group on $2k$ elements $S_{2k}$ \cite{fisher2022random}.  A drastic simplification in the limit of large, but finite, on-site Hilbert space dimension $q$ occurs \cite{zhou2019emergent}, whereby $\langle\left[\Tr\rho_{A}(t)^{2}\right]^{k}\rangle$ may be regarded as $k$ copies of an Ising magnet, with weak inter-replica interactions at each spacetime point where a Haar-random unitary gate has been applied.  The intra-replica interactions for each Ising magnet are described by the statistical mechanical rules presented in Sec. \ref{subsec:stat_mech}. The inter-replica interactions are known to be attractive, and vanish in the limit that $q$ is strictly infinite \cite{zhou2019emergent}.  As already derived in \ref{subsec:stat_mech}, the boundary dissipation acts as an Ising symmetry-breaking field, giving rise to a boundary potential for the Ising domain wall {within} each replica.  

\begin{figure}
    \includegraphics[width=.7\linewidth]{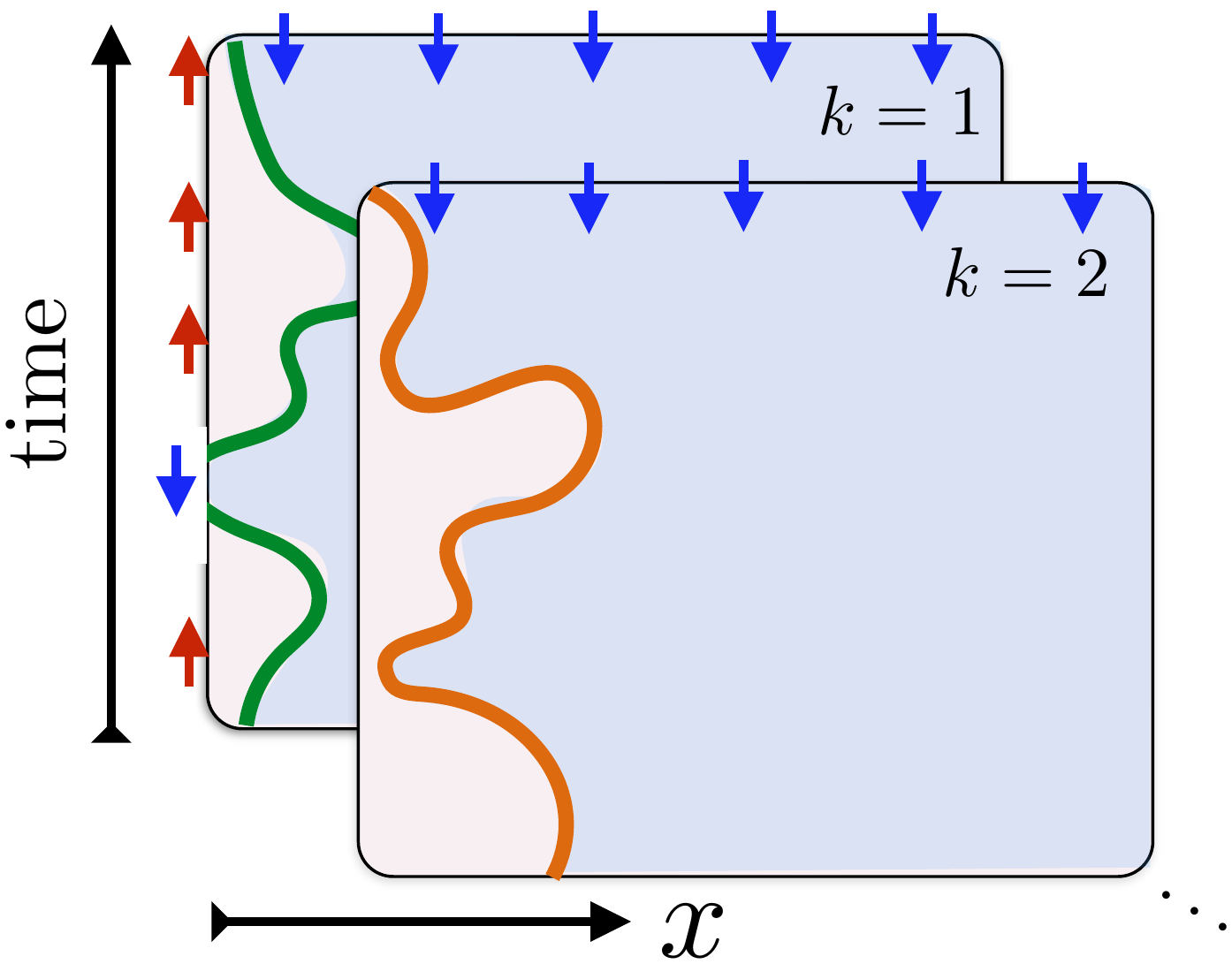} 
    \caption{The Haar-averaged R\'{e}nyi mutual information between the reference qudit(s) and the system, $\langle I^{(2)}_{A,R}(t)\rangle$ is described in the large-$q$ limit, by the $k$ Ising domain walls in the presence of attractive, inter-replica interactions, and an attractive interface within each replica, in the limit $k\rightarrow 0$.  This is described by the path integral in Eq. (\ref{eq:action}). }
    \label{fig:replica_trick}
\end{figure}

The replica limit of the resulting theory may thus be regarded as the description of a directed path in a random environment \cite{HuseHenley1985,kardar1987scaling}, restricted to the half-line $x \ge 0$, and in the presence of a potential near this boundary, due to the dissipation.  The path integral for this problem for a given realization of the disorder is formally given by
\begin{align}
Z[V] = \int\,Dx(\tau)\,e^{-S[x,V]}
\end{align}
where
\begin{align}\label{eq:action}
    S[x,V] \equiv \int d\tau\left[ \frac{1}{2}\left(\frac{dx}{d\tau}\right)^{2} + V[x,\tau] - u\,\delta[x]\right].
\end{align}
Here $x(\tau)$ is the coordinate of the path at time $\tau$. The random potential in the bulk $V[x,\tau]$ is taken to have zero mean, and is short-range-correlated in spacetime, e.g. we may take the potential to be delta-function-correlated as $\overline{V[x,\tau]V[x',\tau']} = \sigma^{2}\delta(x-x')\delta(\tau-\tau')$, where $\overline{\cdots}$ denotes an average over the probability distribution for the disorder.  The statistical mechanics of the replicated theory $\overline{Z^{k}}$ thus describes $k$ interacting paths in the presence of a boundary potential, and thus resembles that of the Haar-averaged quantities $\langle\left[\Tr\rho_{A}(t)^{2}\right]^{k}\rangle$, $\langle\left[\Tr\rho_{A\cup R}(t)^{2}\right]^{k}\rangle$ in the limit of large, but finite, $q$.  A schematic depiction of this replicated theory is shown in Fig. \ref{fig:replica_trick}.

The weak inter-replica interactions are known to be a relevant perturbation at the critical point describing the pinning of a single Ising domain wall \cite{Kardar1985_Depinning}.  Remarkably, the new critical point describing the pinning/de-pinning of a directed polymer to an interface, has been understood exactly \cite{Kardar1985_Depinning} by Bethe ansatz techniques.  The characteristic wandering length of the polymer transverse to the interface diverges with an exponent $\nu_{\perp} = 2$ on approaching the phase transition from the localized phase, while the divergence of the specific heat is characterized by the exponent $\alpha = 0$. For time-independent dissipation (e.g. the depolarizing channel is applied identically at the boundary at each time of the quantum circuit evolution), we thus expect the coding transition to be in the universality class of this de-pinning phase transition for a directed polymer.

In contrast, if the boundary dissipation varies randomly in time - as was studied in Sec. \ref{subsec: CliffordMI} - then the nature of the phase transition is not completely understood.   This problem corresponds to having an imaginary-time-dependent boundary potential $u(\tau) = u_{0} + v(\tau)$ in (\ref{eq:action}), where $v(\tau)$ has zero mean and is short-range-correlated in spacetime; for simplicity, we take $\overline{\overline{v(\tau_{1})v(\tau_{2})}} = \mu^{2}\delta(\tau_{1}-\tau_{2})$, with  $\overline{\overline{\cdots}}$ denoting the average over the distribution for $v(\tau)$.  

We may study the relevance of randomness in this boundary potential at the de-pinning transition.  Here, the action is invariant under coarse-graining and re-scaling $\tau' = \tau/b^{z}$, and $x' \equiv x/b$ where $z$ is the dynamical critical exponent at the phase transition.  Under this transformation, the random boundary potential becomes $\int d\tau\,v(\tau)\delta[x] \longrightarrow b^{z-1}\int\,d\tau'\,v(b^{z}\tau')\delta[x']$, so that we identify $v'(\tau') \equiv b^{z-1}v(b^{z}\tau')$ as the renormalized potential in the coarse-grained theory.  The correlations of the renormalized potential are thus
\begin{align}\label{eq:disorder_scaling}
    \overline{\overline{v'(\tau'_{1})v'(\tau'_{2})}} = \mu^{2}b^{z-2}\delta(\tau'_{1}-\tau'_{2})
\end{align}
Therefore, the strength of the disorder decreases under renormalization when $z < 2$.  It has been conjectured \cite{lipowsky1986wetting} that $z = 3/2$ at the pinning transition for the directed polymer, so that the randomness in the boundary potential should be irrelevant by Eq. (\ref{eq:disorder_scaling}), so that the same fixed-point describing the de-pinning of a directed polymer studied in Ref. \cite{Kardar1985_Depinning} should describe the resulting transition in the presence of randomness. 

We are, however, unaware of the correctness of this result in Ref. \cite{lipowsky1986wetting} for the dynamical exponent.  The numerical studies presented in Sec. \ref{subsec: CliffordMI} further suggest that $\nu_{\parallel} = 2$ (as opposed to $\nu_{\parallel} = z \nu_{\perp} = 3$, which is what would be predicted on the basis of $z = 3/2$ and $\nu_{\perp} = 2$), though more extensive numerical studies are required to pin down the nature of this transition.  We note, for completeness, that Eq. (\ref{eq:disorder_scaling}) suggests that the random boundary potential is a marginal perturbation exactly at the de-pinning phase transition for the Ising domain wall (which has $z=2$  \cite{chalker1981pinning}).  A Wilsonian renormalization-group calculation to higher order further suggests that the disorder is marginally \emph{relevant} \cite{SV_MPAF_Unpub}. The nature of the resulting critical point is not understood, and deserves further investigation.  %A fully consistent picture would thus require that the resulting critical point would also be unstable to inter-replica interactions, which would then 

\subsection{Perfect information protection using scrambling}~\label{subsec:logscrambling}

In the low-dissipation phase of the coding transition, quantum information is only partially protected. One would expect that the information protection can be improved by first scrambling the information with unitary gates, which can effectively act like a random encoding, before the dissipation is turned on; we refer to this as a ``pre-scrambling" step.
Here we argue that for fixed system size $L$ and dissipation strength $p$, scrambling the initially local quantum information via a random unitary circuit of logarithmic depth $ \tscr = k\log L$ for some sufficiently large $k$, can lead to perfect protection of quantum information within the system, up to times of order $\T\sim L/p$. For a pre-scrambling step with a fixed depth $t_{\mathrm{scr}} = k\log L$ and for low $k$, we can observe the coding transition by tuning the dissipation strength $p$.  The coding transition will now be manifest in a step-function-like behavior of the mutual information $I_{A,R}$ across the transition due to the perfect preservation of information for sufficiently low dissipation.

 \begin{figure}
    \includegraphics[width=.83\linewidth]{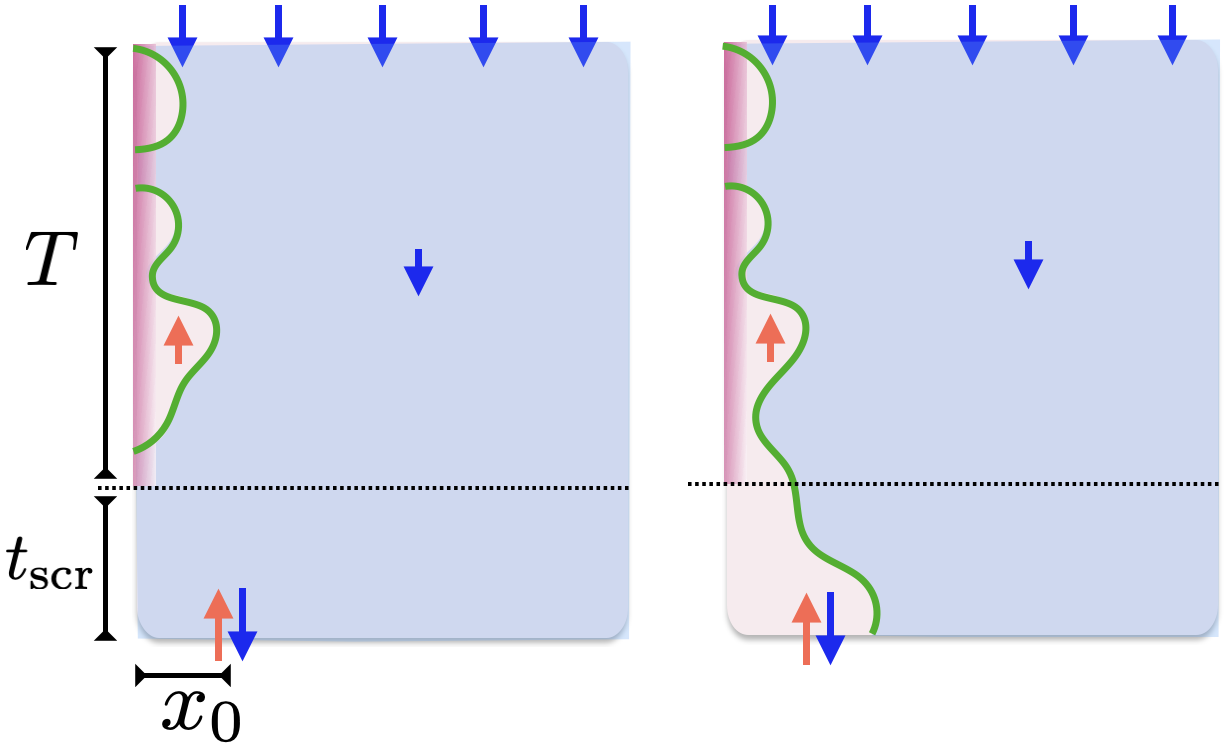} 
    \caption{The behavior of the Ising domain wall in the presence of a pre-scrambling step, whereby the initially local quantum information is evolved by a quantum circuit of depth $t_{\mathrm{scr}}$.  We consider propagation of the domain wall backwards in time, with respect to the arrow of time in the quantum circuit.  In this picture, trajectories of the domain wall which are survive in the bulk into the pre-scrambling step (right) are exponentially suppressed relative to trajectories which are annihilated at the boundary beforehand (left). }
    \label{fig:log_scrambling_dw}
\end{figure}

To gain some intuition for this result, we again consider the statistical mechanics of the Ising domain wall. As before, the domain wall is naturally thought of as propagating in a direction which is opposite to the arrow of time in the quantum circuit evolution.  The domain wall thus propagates through $T$ timesteps of the circuit involving boundary dissipation, and then encounters the pre-scrambling step where the dissipation is absent.  This corresponds to free evolution of the domain wall without the symmetry-breaking field at the boundary. When this field at the boundary is turned off, trajectories of the domain wall which have already been annihilated at the boundary -- such as the one shown in the left panel of Fig. \ref{fig:log_scrambling_dw} -- do not cost additional weights in the partition sum. On the other hand, ``surviving" domain wall trajectories in the bulk -- such as the one shown in the right panel of Fig. \ref{fig:log_scrambling_dw} -- incur a weight of $q/(q^2+1)$ at each time step. {Thus the weights of the bulk trajectories of the domain wall are exponentially suppressed in time relative to trajectories terminating at the boundary.

 Let $Z_{a}(t,T)$ be the partition function for the Ising domain wall, after the $T$ timesteps of the dynamics with dissipation have taken place, followed by an additional $t$ timesteps of pre-scrambling, and so that the domain wall has been annihilated at the boundary of the system.  In contrast, let $Z_{b}(t,T)$ be the partition function for the Ising domain wall to ``survive" in the bulk of the system after the same evolution. To determine the behavior of the annealed mutual information, we wish to determine the probability that the domain wall ends at position $x\ge x_{0}$ after another $t$ steps of the dissipation-free evolution, as per Eq. (\ref{eq:annealed_MI}), where $x_{0}$ is the location of the entangled reference qubit of quantum information.  For simplicity of presentation, we take $x_{0}$ to be at the boundary of the qubit chain, so that this probability $P(t,T)$ is  
 \begin{align}
 P(t,T) &= \frac{ Z_{b}(t,T)}{Z_{a}(t,T) + Z_{b}(t,T)}
 \end{align}
 
To make progress, we note that since the ``surviving" trajectories contributing to $Z_{b}(t,T)$ are exponentially suppressed in time, we may write that $Z_{b}(t,T) = Z_{b}(0,T)e^{-\gamma t}$, where $\gamma$ is a phenomenological decay rate which will be a function of the local Hilbert space dimension, and the dissipation strength.  We further approximate the partition sum $Z_{a}(t,T)$ by its value before the pre-scrambling step, so that $Z_{a}(t,T) = Z_{a}(0,T)$. With these approximations, we may write
 \begin{align}
P(t,T) &= \frac{P(0,T)}{P(0,T) + [1 - P(0,T)]e^{\gamma t}}
\end{align}

 The annealed mutual information is now obtained from Eq. (\ref{eq:annealed_MI}).  At sufficiently long times, so that $P(t,T)\ll 1$, we thus find that the mutual information deviates from its maximal value by 
 \begin{align}
2 - I^{(\mathrm{ann})}_{A,R}(t) = \frac{q^{2} - 1}{q}\cdot\frac{P(0,T)}{P(0,T) + [1 - P(0,T)]e^{\gamma t}}
 \end{align}
 In the pinned phase of the domain wall, we expect $P(0,T)$ is exponentially small in the number of timesteps $T$.  In contrast, in the de-pinned phase, the probability that the domain wall has been annihilated at the interface decays as a power-law in time due to the diffusive nature of the Ising domain wall, so that $P(0,T) = 1 - O(T^{-a})$, with $a$ a constant.  For fixed $T$, we thus find that for a sufficiently long pre-scrambling time $t$, the mutual information deviates from its maximal value as
 \begin{align}
	2-I^{(\mathrm{ann})}_{A,R}(t) \sim \begin{cases}
		e^{-\gamma t} & p<p_c \\
		\T^a e^{-\gamma t} & p>p_c
	\end{cases}.
\end{align}
 Evaluating this expression at the scrambling time $ \tscr=k\log L $ yields 
 \begin{align}
	2-\MIann(t) \sim \begin{cases}
		L^{-\gamma k} & p<p_c \\
		L^{a-\gamma k} & p>p_c 
	\end{cases}.\label{eq: annealed MI log scram}
\end{align}

\begin{figure}
$\begin{array}{c}
    \includegraphics[width=\linewidth]{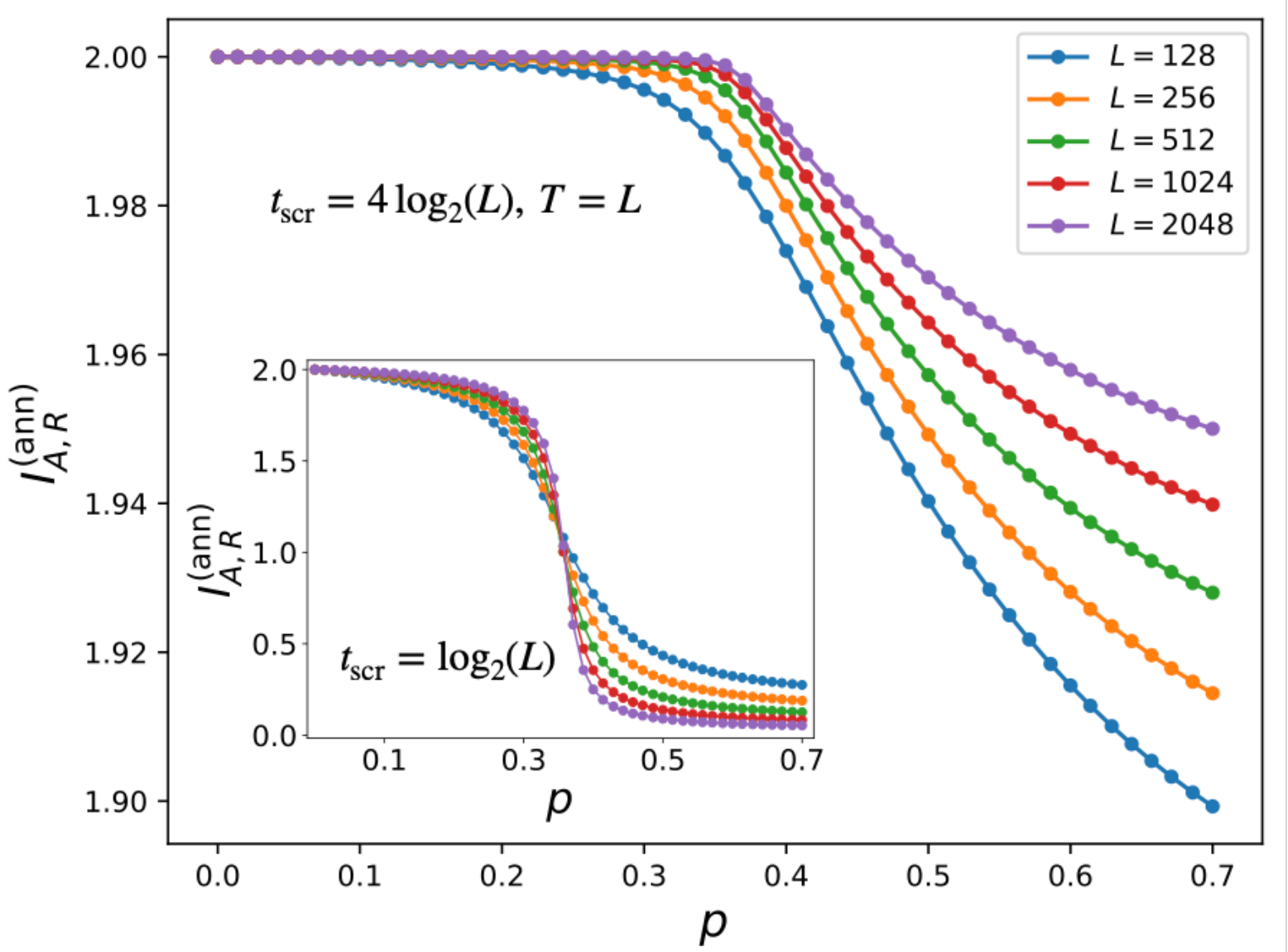}\\
    \text{(a)}\\   
    \includegraphics[width=\linewidth]{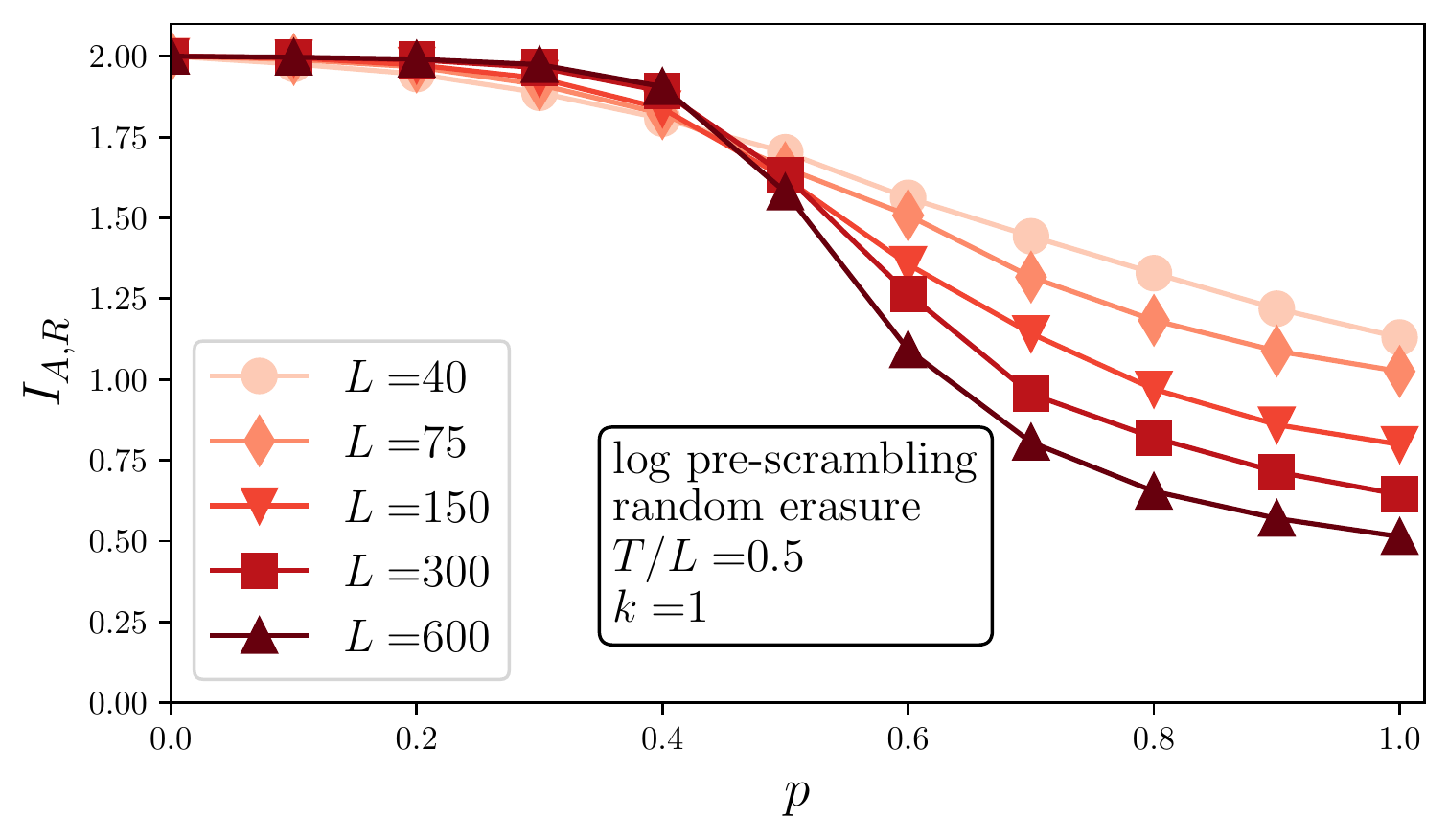}\\
    \text{(b)}
\end{array}$
    \caption{Coding transition with logarithmic-depth pre-scrambling. In (a), $\MIann$ vs $p$ is plotted with a pre-scrambling circuit of depth $\tscr \sim \log L$.  The subsequent evolution with dissipation proceeds for a total number of timesteps $\T=L$. The main plot is for $\tscr=4\log_2(L)$. The annealed mutual information approaches the maximum value as $L$ is increased indicating that logarithmic-depth encoding is enough to protect the information against boundary dissipation. \textit{Inset} shows the plot for $\tscr=\log_2(L)$ with $\MIann$ going through a transition with respect to $p$. The results agree with eq.~\eqref{eq: annealed MI log scram} derived in the main text.  In (b), the  mutual information, as calculated in Clifford dynamics, for dynamics with pre-scrambling of depth $\tscr=\log_2(L)$, plotted as a function of dissipation strength $p$. Boundary dissipation is realized as a random erasure channel, and $T/L = 1/2$ is kept fixed for different system sizes. The mutual information reveals a phase transition, with the critical point appearing as a crossing point of the data for different system sizes. }
    \label{fig:logscramble}
\end{figure}

The above calculation implies that for $ \tscr = k\log L $, with $ k $ large enough, quantum information is perfectly preserved. Logarithmic scrambling is enough to protect the information against noise. For low values of $ k $, the mutual information can exhibit different behavior depending on whether $ a-\gamma k $ is positive or negative. 
We show the results obtained from studying the annealed MI numerically in Fig.~\ref{fig:logscramble}a, and find good agreements with the considerations above.

We now turn to the simulation of Clifford quantum circuit dynamics.  To explore how logarithmic pre-scrambling affects the coding transition induced by a single boundary,  we modify the circuit protocol   to include a unitary, non-dissipative pre-scrambling step, with pre-scrambling time scaling logarithmically with system size, $\tscr =k \log L$, before applying the dissipative dynamics for time $T$. We then approach the thermodynamic limit by increasing $\T$ and $L$, while keeping the aspect ratio $T/L<1$ fixed. In accordance with the insights gained above from the annealed Haar average, we find a phase transition for $k=1$ as a function of $p$ between a phase retaining information between the input and output of the circuit, and a phase with all information destroyed by dissipation, as shown in Fig.~\ref{fig:logscramble}b. The critical properties are different from the case without pre-scrambling discussed in the previous subsection, and, as predicted by the annealed model, the critical point is signaled by a crossing point in the mutual information obtained for different system sizes. We find a similar coding transition for $k\leq k_{\rm max}$, with  $k_{\rm max}\sim O(1)$. For even larger values of $k$, the mutual information remains maximal for all values of $p$.

\section{Coding transition on the approach to thermalization}\label{sec:disentangling_transition}
In the previous section, we studied systems of size $L$ with dissipation acting near the left boundary in the regime $T \lesssim L$ so that the right boundary did not play a role in the dynamics. More precisely, as long as $L/T$ remains larger than the velocity of the entanglement domain wall, which is less than the lightcone velocity in the quantum circuit, the coding transition can be understood as a depinning transition of the domain wall, such that for noise rate $p$ below the critical value $p_c$ some amount of information survives. 

In this section, we study what happens when the dynamics in the coding phase extend for even longer periods of time, and show that the surviving information will eventually be lost to the environment as the system completely thermalizes.  
We may understand this result by considering the dynamics of the Ising domain wall, which describes the behavior of the annealed mutual information.  For sufficiently large $\T/L$ the domain wall
will escape and get annihilated at the right boundary. Thus using eq. \eqref{eq:annealed_MI} $\MIann$ becomes zero and the information gets leaked to the environment. Intuitively speaking, the system gets entangled with $pT$ number of environment qubits, and when $pT \gtrsim L$ the system
gets maximally entangled with the environment and become thermalized. By the
monogamy of entanglement, the reference qudits can no longer be entangled with the system but are lost to the environment. Therefore for large $\T /L$ there is a transition with
respect to the dissipation strength $p$, and the location of
the critical point scales as $\pd \sim \T /L$; for $p>\pd$ the information gets completely entangled with the environment. 
This transition is also visible with respect to $\T$ and fixed dissipation strength $p$.

We study this coding transition by performing $\tscr = L$ steps of pre-scrambling before turning on the noise. As explained in the previous section, linear pre-scrambling perfectly protects the information for all strengths of dissipation, and when $T/L$ is sufficiently small. This pre-scrambling step has the effect of making the transition appear as a ``step function" in the mutual information $I_{A,R}$ as a function of dissipation strength. Indeed, $ I_{A,R}^{(\mathrm{ann})}(\T) $ vs $ p $ for $ \T/L=4 $ in Haar random circuit in Fig. \ref{Fig. high t transition} shows such a behavior, and appears to be a scaling function of $(p-p_d)L$ (see inset). 

\begin{figure}
	\includegraphics[width=\linewidth]{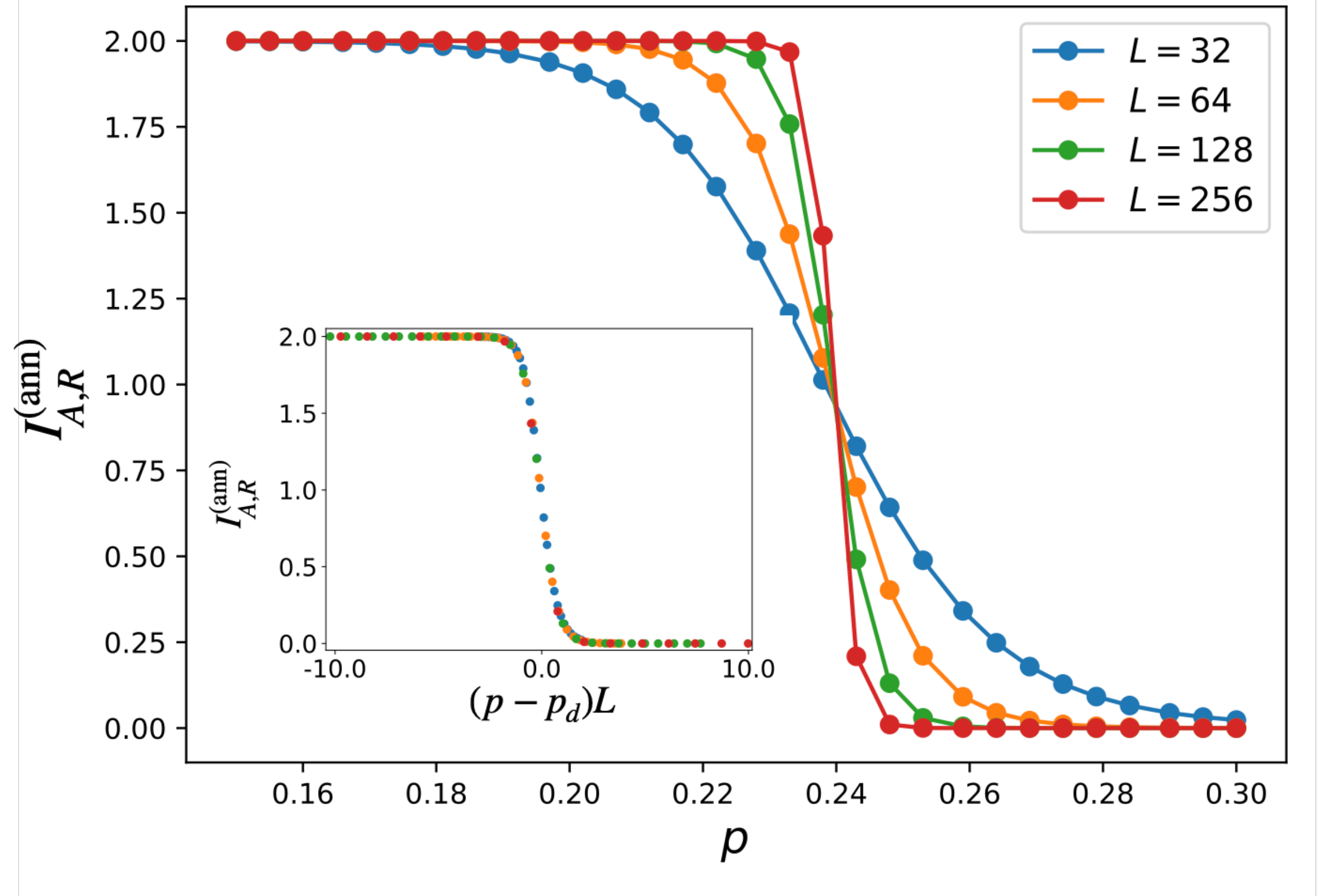}
	\caption{Plot of $ \MI_{A,R}^{(\mathrm{ann})} $ in Haar random circuits. $ \T/L=4 $ and $ \tscr = L $. \textit{Inset.} The data collapse to a single curve as a function of $(p-\pd)L$.} \label{Fig. high t transition} \label{fig: disentangling transition annealed}
\end{figure}

\subsection{Numerical Study}\label{subsec:disentangling_numerics} 

We also verify the above transition in the Clifford circuit setting introduced in the previous section. Here, after initializing the Bell pair at the left boundary of the chain, we run a pre-scrambling step linear in system size, $\tscr = L$, followed by the dissipative dynamics applied for time $\T$. As before, we examine the finite size scaling by increasing $\T$ and $L$, while keeping $\T/L>1$ fixed.
As already discussed in the annealed framework, we find a phase transition for large enough aspect ratio $\T/L>1$. In Fig.~\ref{fig:I_largeT}a, we plot the mutual information between the reference qubit and the output of the circuit as a function of $p$ for different system sizes $L$, using a pre-scrambling time $\tscr=L$ and aspect ratio $\T/L=4$. In perfect agreement with the annealed picture, the mutual information curve approaches a step function in the thermodynamic limit, confirming a phase transition between a phase with all the information protected, and a phase with all information destroyed.

We find a good scaling collapse with the scaling function depending on $(p-p_d)L^{1/2}$, see Fig.~\ref{fig:I_largeT}b. The form of the scaling function differs from the annealed result. This deviation can be understood by noting that for the annealed case we applied a deterministic boundary depolarization channel, Eq.~\eqref{eq:depolarize} whereas the dissipation in the Clifford circuit is applied at random time steps, and this disorder may change the properties of the transition.
Indeed, the effect of randomness in the dissipation channel can be studied by introducing disorder into the annealed model and applying channel ~\eqref{eq:depolarize} at random times which leads to scaling function depending on $(p-p_d)L^{1/2}$ (data not shown), in perfect agreement with the Clifford circuit results. {The discrepancy between the factor of $L$ and $L^{1/2}$ can be understood as follows. With randomness, the number of environment qubits entangled with the system increase linearly with $\T$ but has fluctuations of order $\sqrt{T}$. This results in the critical point fluctuating as $\delta p/\sqrt{T}$ leading to $(p-p_d)L^{1/2}$ dependence of the mutual information.}

\begin{figure}
    \centering
    \includegraphics[width=\linewidth]{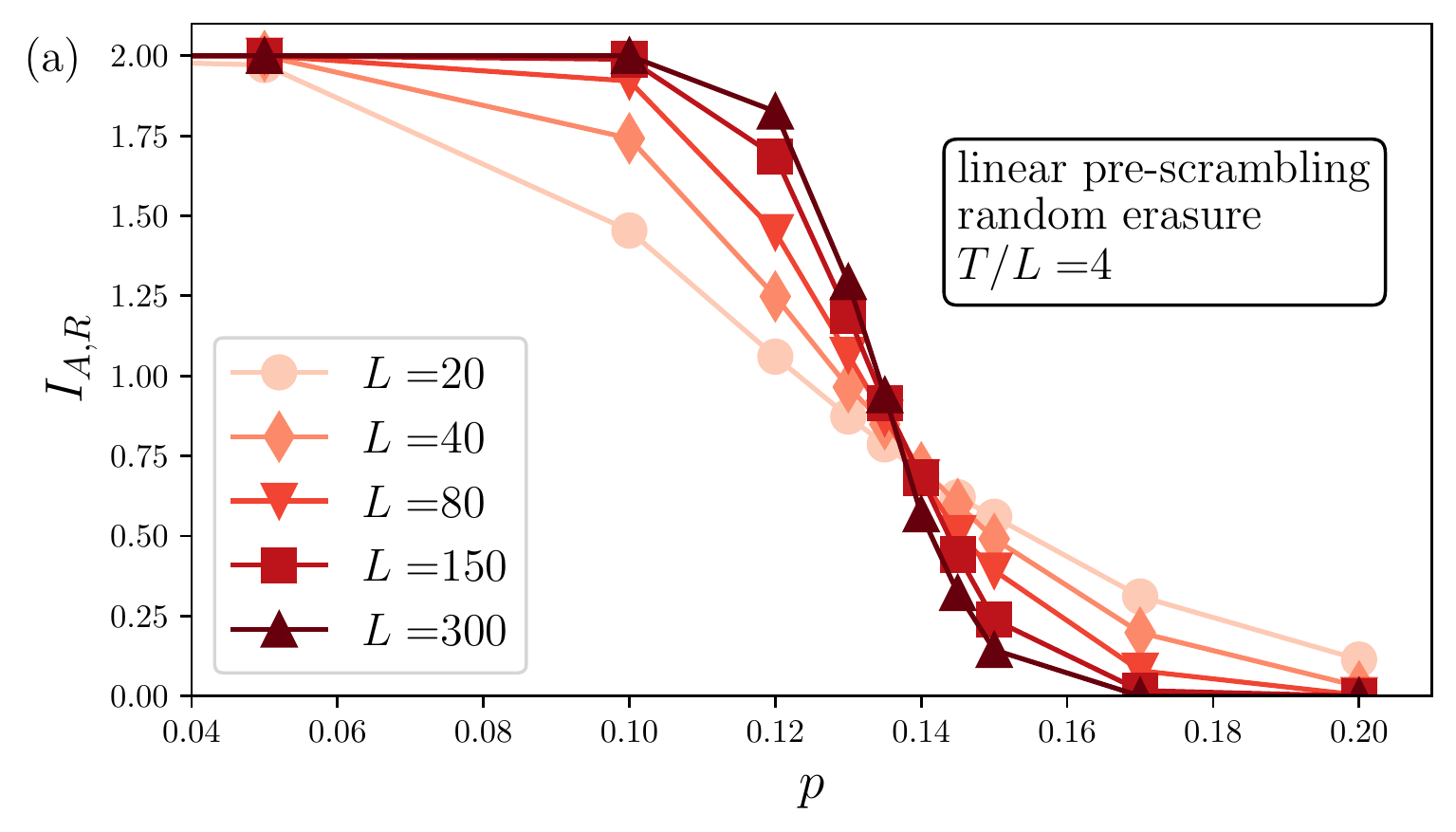} \\
    \includegraphics[width=\linewidth]{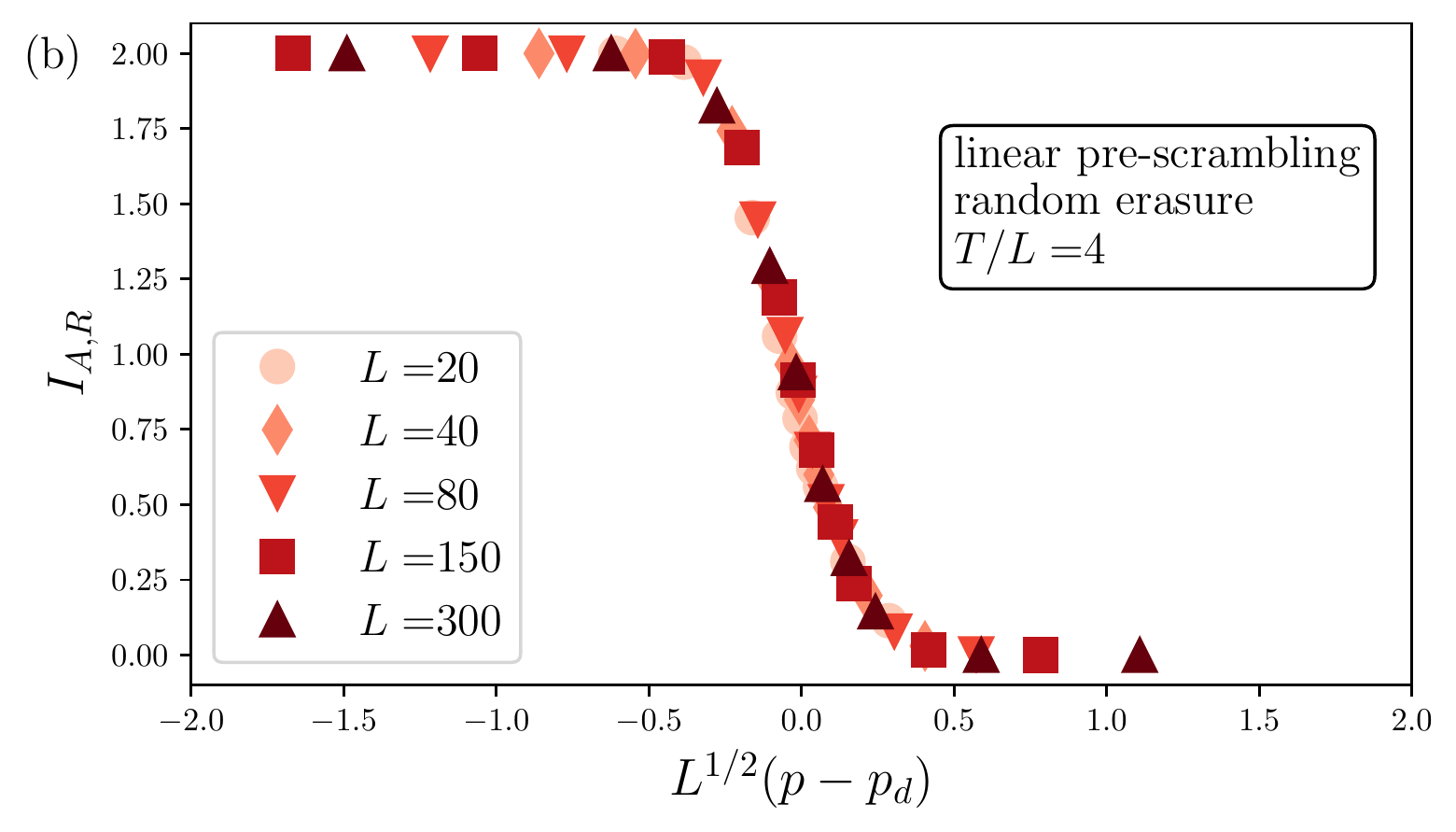} 
    \caption{Coding transition upon approaching thermalization. (a) Mutual information between the input and the output of the circuit shown as a function of dissipation strength $p$, converging towards a step function in the thermodynamic limit. Pre-scrambling time is set to $\tscr=L$, followed by dissipative dynamics for time $\T$, with $\T/L=4$ fixed. (b) Data collapse as a function of $(p-p_d)L^{1/2}$, with the critical point $p_d=0.136$ corresponding to the crossing point of finite size data.}
    \label{fig:I_largeT}
\end{figure}

\subsection{Nature of the Phase Transition}\label{subsec:disent_discussion} 

We end this section by discussing the nature of the transition explored above. We argue below that coding transition in this regime is a first-order phase transition. 

To begin with, let us consider the large qudit limit such that $1/q \ll (1-p)^2$. The partition function in the annealed picture contains contributions coming from all possible trajectories of the domain wall. The contribution at time $t$ from trajectories having the domain wall at $n_{DW}$ number of time steps is of order $(1/q)^{n_{DW}} ((1-p)^2)^{t-n_{DW}}$. The entropic factor, due to there being more configurations with the domain wall as opposed to without it, can only renormalize the $1/q$ factor. Thus the partition function is dominated by the term having no domain wall at any point of time, $(1-p)^{2t}$. However, for $(1-p)^{2t} > (1/q)^L $, it is preferable for the domain wall to go all the way to the right boundary and get annihilated there. Thus at $t_c \sim \frac{\log 1/q}{\log (1-p)} L $ the nature of the domain wall changes discontinuously from being stuck at the noisy boundary to getting annihilated at the un-noisy boundary indicating a first-order transition.
The finite $q$ corrections to the above picture only act as thermal fluctuations which causes the domain wall to have some excursions inside the bulk. The contributions from these excursions will be sub-leading and we expect the transition to remain first-order. {Note that similar time scales were also identified in~\cite{Li_Sang_Hsieh_2023} for the system to become perfectly thermalized in the presence of noise.}

As in the standard theory of first-order phase transition, the two boundaries correspond to the two local minima for the domain wall and the system discontinuously jumps from one to another. The mutual information then is a function of the probability that the system is in one of the two minima (see eq.~\eqref{eq:annealed_MI}). Since the free energy is extensive, the probability of being in a particular minimum scales as a function of $\delta g V$ where $\delta g$ is the tuning parameter for the transition and $V$ is the total volume of the system. In our case, the volume is equal to $\T$. This explains the observed finite-size collapse as a function of $(p-\pd)\T$ in Fig.~\ref{fig: disentangling transition annealed}.

\section{Encoding at a Finite Rate} \label{sec:finite_code_rate}

So far we looked into the dynamics of a single bell pair localized near the noisy boundary. But it is equally interesting to understand the effects of the noise when we have an extensive number of Bell pairs in the initial state. We denote the code rate, defined as the fraction of the system's qubits entangled in Bell pairs, by $C=N_R/L$ where $N_R$ is the total number of Bell pairs. For the purpose of this section, we will consider code density $C=1/2$ but we believe that the qualitative results should not change for different values of $C$ as long as $C$ is not close to $1$. To make the final results independent of the distribution of the Bell pairs at the initial time we will perform random encoding by performing unitary scrambling for time $\tscr=L$.

We plot the annealed mutual information between the input and output, $\MI_{A,R}^{(\mathrm{ann})}$, in Fig. \ref{fig: extensive rate} as a function of the dissipation strength for $\T=7L$. We find two threshold values for the noise rate, $p_{th,1}, p_{th,2}$. For $p<p_{th,1}$, the information is perfectly protected and $\MIann$ is equal to the maximal value $2CL$. For $p_{th,1}<p<p_{th,2}$, the information starts leaking to the environment but still a finite density of it remains in the system. Finally when $p>p_{th,2}$ the information is completely leaked to the environment. Note that the values of $p_{th}$ change with the ratio $\T/L$.

 Similarly to the strategy followed in the previous sections, we verify these predictions by performing numerical simulations in Clifford quantum random circuits. We show the density of the mutual information between the output of the circuit $A$ and the reference qubits, $I_{A,R}/N_R$, with $N_R=L/2$ denoting the number of input Bell pairs, as a function of dissipation strength $p$ in Fig.~\ref{fig:I_finiterate}, for different system sizes $L$ with $T/L=4$ fixed. As noted above, here we applied a linear unitary pre-scrambling step for time $\tscr=L$, before the onset of the noisy dynamics, such that the results do not depend on the spatial distribution of the Bell pairs in the initial state. We find a phase with perfectly protected information for small enough dissipation strength $p$, followed by a crossover region with a finite density of preserved coherent information decreasing continuously with $p$, eventually decaying to zero for large $p$. 

\begin{figure}
    $\begin{array}{c}  
         \includegraphics[width=\linewidth]{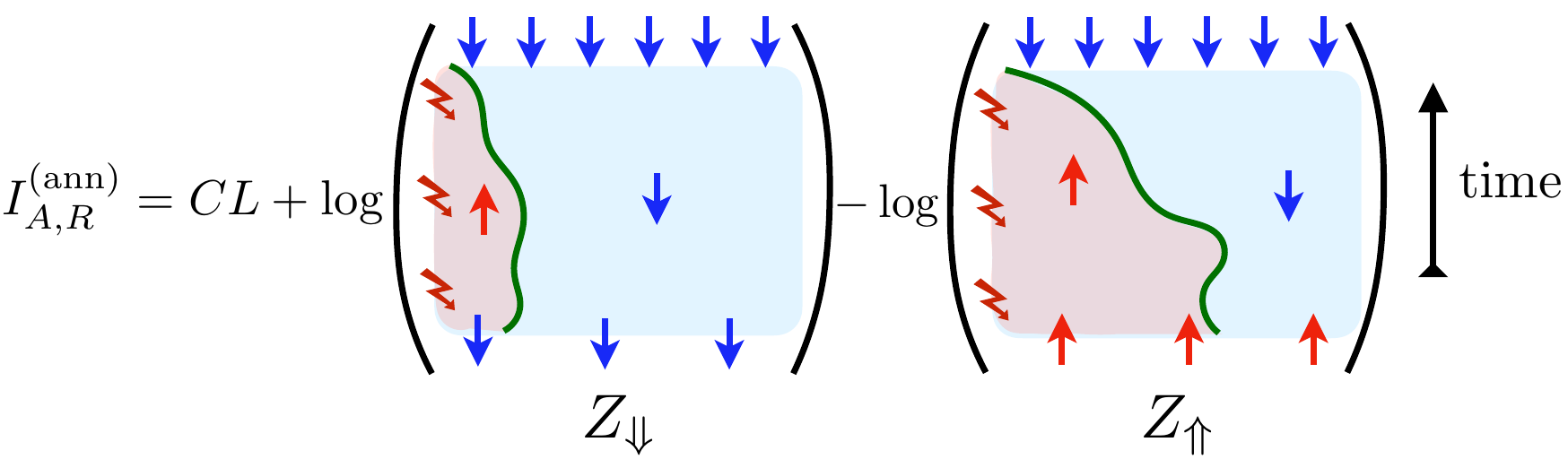} \\
          \\
          \includegraphics[width=.83\linewidth]{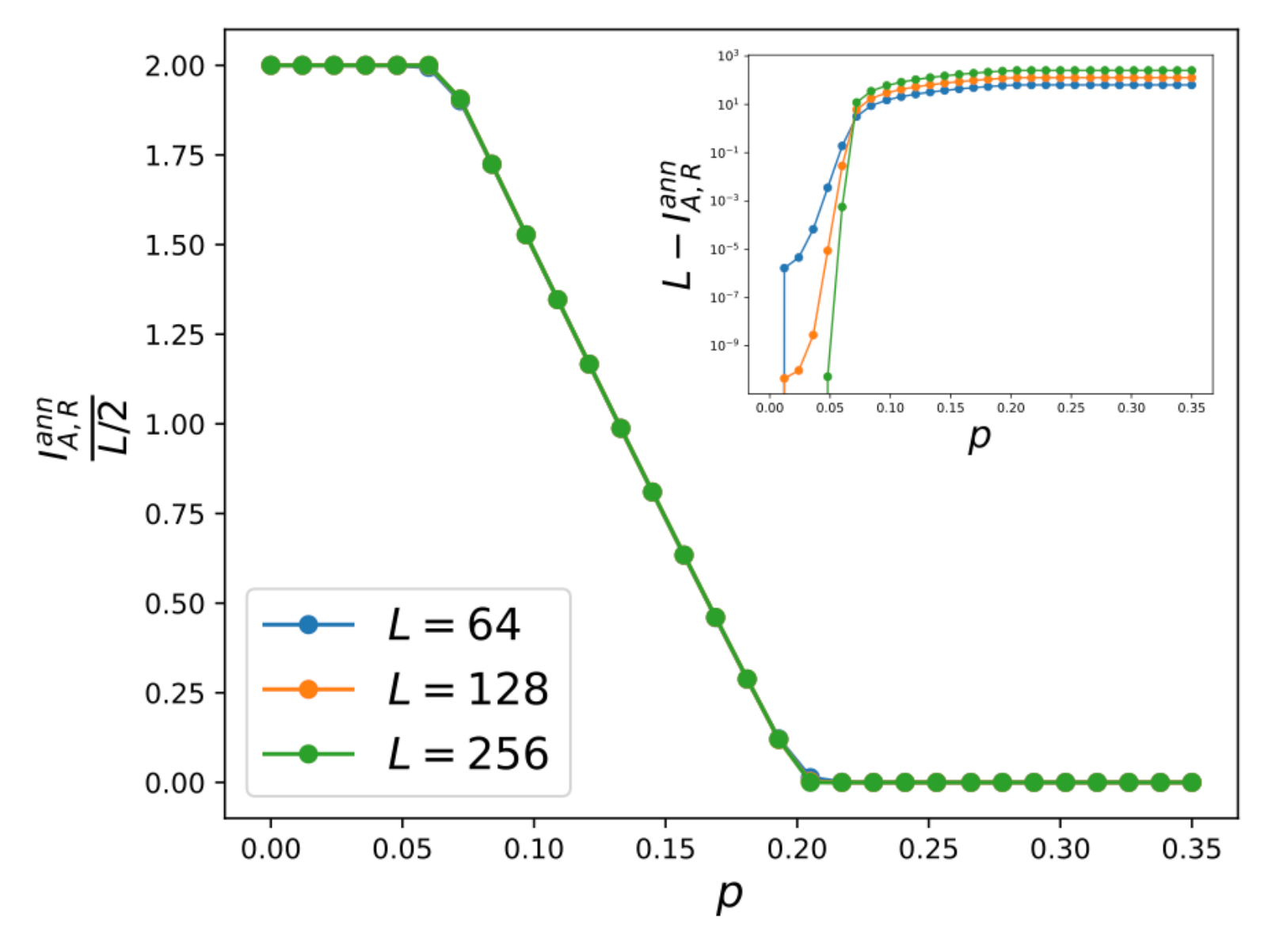} 
    \end{array}$
    \caption{\textit{Top.} Schematic representation of the statistical mechanics of the Ising domain wall in the calculation of the annealed mutual information, when coding at a finite rate. Typical domain wall trajectories when $p_{th,1}<p<p_{th,2}$ are shown. In  $Z_\Downarrow$ the domain wall remains localized whereas it is delocalized for $Z_\Uparrow$, as explained in the text. \textit{Bottom.} Plot of the annealed mutual information between Bell pairs entangled with the system's qubits at alternate sites ($C=1/2$) and the system. The Bell pairs are scrambled by a unitary circuit for time $\tscr=L$. The system is evolved in presence of the boundary dissipation for time $\T=7L$. We find that for $p<p_{th}^1\approx 0.06$, full information is preserved, while for $p_{th}^1<p<p_{th}^2\approx 0.2$, a finite density of information is protected. The threshold values decrease as $\T$ is increased. \textit{Inset.} For low $p<p_{th}^1$ there is no information loss even for $\T=7L$, that is, the difference between $\MIann$ and the maximum value $L$ goes to zero with system size. Thus all Bell pairs can be perfectly recovered by a recovery operation acting on the system.} \label{fig: extensive rate}
\end{figure}

\begin{figure}
    \centering
    \includegraphics[width=\linewidth]{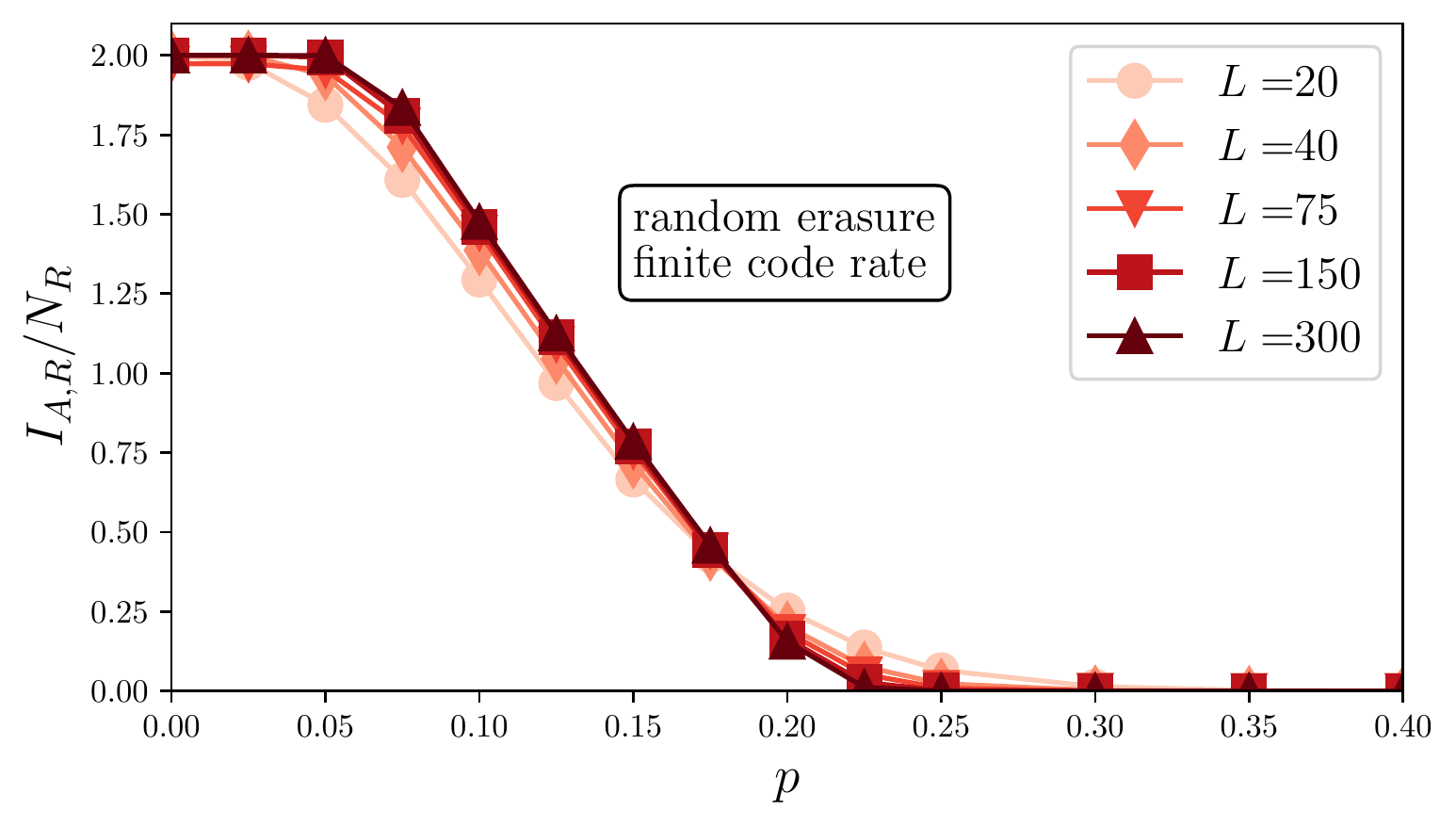} \\
    \caption{Coding transition for finite code rate. Density of mutual information between the output of the circuit and the reference qubits shown as a function of dissipation strength $p$, for fixed evolution time $T/L=4$ and number of initial Bell pairs $N_R=L/2$. Pre-scrambling time is $\tscr=L$, followed by noisy dynamics with a random boundary erasure channel. The information density is perfectly protected for weak enough dissipation $p$, then decays continuously  towards zero with $p$ in a crossover region, with all information leaked to the environment for $p$ large enough.}
    \label{fig:I_finiterate}
\end{figure}

To understand this behavior we again resort to the statistical mechanics of the Ising domain wall. This model for a finite code rate differs importantly from that of the model when an $O(1)$ amount of quantum information is encoded.  In the case of finite coding rate there are an extensive number of Ising spins at the top boundary {whose state is fixed by the boundary conditions, though the bulk dynamics of the domain wall remain the same}. This leads to an exponential amplification of the trajectories that minimize the number of domain walls at the top boundary (note that these domain walls at the boundary are different than the Ising domain wall performing random walk in the bulk). As shown at top of Fig. \ref{fig: extensive rate}, the annealed $\MI$ is given by
\begin{align}
    \MIann = CL + \log\left(\frac{Z_{\Downarrow}}{Z_\Uparrow}\right) \label{eq: MI extensive}
\end{align}
where $Z_{\Downarrow}, Z_{\Uparrow}$, are the partition function of the statistical mechanics model with down and up spins respectively at the locations of the encoded Bell pairs; the log is in the base of $q$. As discussed in Sec.~\ref{sec:disentangling_transition} the domain wall discontinuously changes from being at the left boundary to being at the right boundary. To a good approximation, we can thus only keep these two trajectories in the partition function. For clarity of the expressions we also introduce $\Tilde{p}\equiv 1-p$. The partition functions $Z_{\Downarrow},Z_{\Uparrow}$ can thus be written as \begin{align}
    Z_{\Downarrow} \approx \Tilde{p}^{2\T}q^{2CL} + \left(\frac{1}{q}\right)^{L} q^{CL} \\
    Z_{\Uparrow} \approx \Tilde{p}^{2\T}q^{CL} + \left(\frac{1}{q}\right)^{L} q^{2CL}
\end{align}
Putting the above expression in eq.~\eqref{eq: MI extensive} and identifying the threshold values to be $1-p_{th,1} \sim q^{-(1-C)L/(2\T)}, 1-p_{th,2} \sim q^{-(1+C)L/(2\T)}$, we get\begin{align}
    \MIann \approx \begin{cases}
        2CL & p < p_{th,1} \\
        2CL - 2\T\log \left(\frac{1-p_{th,1}} {1-p} \right) & p_{th,1} < p < p_{th,2} \\
        0 & p > p_{th,2}.
    \end{cases} 
\end{align}

{Intuitively, for low $p$ the domain wall remains localized near the noisy boundary and mutual information is maximal. As $p$ is increased, it is easier for the DW in $Z_\Uparrow$ to delocalize compared to $Z_\Downarrow$ as in the former delocalization results in an exponential reduction in the cost associated with having domain walls at the boundary. Thus the critical point at which the DW delocalizes is different for the two boundary conditions resulting in the two thresholds discussed above.}

\section{Summary and Discussion}\label{sec:discuss}

In this work, we studied one-dimensional quantum many-body systems with a noisy boundary. We focused on the dynamics of the information of an initially localized Bell pair near the (noisy) boundary by studying the mutual information $I_{A, R}(t)$ between the inert spin of the Bell pair with the system at later times where $A$ is the system and $R$ is the inert spin. This is also related to the coherent information about the Bell pair remaining in the system~\cite{Schumacher1996,Schumacher2001}.
We find that the chaotic scrambling due to the unitary dynamics is sufficient to protect a part of this information against getting leaked to the environment for noise rate $p<p_c$ and long times $\T\lesssim L/p$ by allowing the information to escape away from the boundary. We further show that a random encoding of the Bell pair via noise-less scrambling dynamics of depth $\mathcal{O}(\log L)$, is sufficient to \textit{perfectly} protect the information for all strengths of the noise upto time $\T\lesssim L/p$. See~\hbox{Fig.~\ref{fig: summary_figure}.b} for a schematic representation of the phase diagram.

In the regime when the total time of evolution $\T \gtrsim L/p$, any remaining information in the system is revealed to the environment and the system go through a first-order coding transition. 
This transition can also be seen as a result of the system approaching thermalization to infinite temperature. We expect this form of coding transition to be present in all noisy channels though in the case of the boundary noise considered here, the timescales associated with the transition increase parametrically with the system size~\cite{Li_Sang_Hsieh_2023}.

We also look at the coding dynamics for finite code rate, that is, when an extensive number $N_R = CL$, with $C<1$, of the system's qubits are entangled in Bell pairs. We find that the code space can be \textit{perfectly} preserved for noise strength below some threshold $p_{th,1}$ and for strength above $p_{th,2}$ the code space is completely destroyed, see Fig.~\ref{fig: extensive rate},~\ref{fig:I_finiterate}. We can also look at the time for which the information stays in the system for a fixed noise rate $p$ and equivalently define two threshold times $\T_{th,1}<\T_{th,2}$ both of which scales linearly with system size.

This work provides new insights into the competition between scrambling and decoherence. Normally, active feedback in the form of error correction is needed to counter the decoherence effects of the noise. However, we present the case of boundary noise where it is possible to have stable quantum error codes (QEC) in presence of generic noise, with the code space dynamically protected by scrambling. Previously such dynamical protection of information was also observed for the special case of dephasing noise which can be unraveled into quantum trajectories corresponding to projective measurements, but there an extensive number of ancilla qubits that act as register for the measurement outcomes are made part of the system ~\cite{Gullans_Huse}. It would be of interest to generalize our results and techniques in the presence of ancilla qubits for cases different from the boundary noise. We leave this for future work.

{Other interesting directions to explore are the presence of similar coding transitions in purely unitary evolution. It seems possible for quantum information to remain confined in part of a system evolving under chaotic unitary dynamics for a long time, and before the system thermalizes. We leave a detailed discussion of this direction to future work~\cite{non_markovian_Unpub}. }

The competition between chaos and decoherence has also been studied in the context of open quantum systems. Previous studies have mostly focussed on level statistics and quantities like spectral form factor, purity, and Loschmidt echo to study the effect of decoherence in chaotic dynamics~\cite{kawabata2022dynamical,Yan_2020,Xu_decohernce_2019,Can2019,Jalabert2001,Karkuszewski2002,Cucchietti2003,Peres1984}. It is an open question to study such probes in our context and whether the coding transitions can also be seen in these quantities. There is also a close relationship between the input-output mutual information and operator spreading (measured via out-of-time-correlators (OTOCs)) in noise-free unitary dynamics~\cite{Hosur2016}. It is interesting to understand how OTOCs in noisy systems are related to the emergent QEC property of the noisy dynamics~\cite{schuster2022operator,Zanardi2021,Yoshida_Yao_2019}. Or more generally, how is the dynamics of information related to the above-mentioned quantities for open quantum systems?

The coding transitions imply protection of the code-space against noise and the potential existence of a decoding protocol that brings back the code-space to its initial state. Such a protocol is notoriously hard for random dynamics having little structure, except in a few special cases like the Preskill-Hayden black hole protocol~\cite{Hayden2007,yoshida2017efficient} or for special types of noises like erasure channel. For Clifford circuits with boundary dissipation considered here, an efficient decoder can probably be constructed for the erasure channel. Another interesting direction in further understanding the error-correcting properties of the coding transitions is to look into the code distance of the resulting code. We leave a detailed study of the decoding protocols and code distance for future studies.

We also find similar coding transitions for bulk defects where noise acts on the same site in the bulk. Protection of quantum information against bulk defects is important for the design of modular quantum computers in which smaller modules of quantum memory/computer are connected together to form a bigger block. In this case, one expects the noise in the gates connecting the two modules to be far greater than the noise in the bulk of the individual modules. Thus the existence of an error-threshold against a bulk defect and the availability of the decoding protocol discussed above gives a fault-tolerant way of building a modular quantum computer.

A possible extension of our work is to study information dynamics in noisy symmetric systems.  The behavior of information in symmetric systems with local charge density in presence of measurements has been shown to be qualitatively different than without symmetry~\cite{Agrawal_2022,Barratt_2022,Barratt_2022_field,Oshima2023}. It is also known that systems with local charge conservation can have charge transport and long-time operator entanglement growth even in the presence of strong dephasing noise~\cite{Wellnitz2022_depahsing_U(1),Cai_2013}. This may potentially lead to a more robust encoding of the information for when the {code-space} is spread across different charge sectors as opposed to being confined to one sector. We leave this for future studies. }

\acknowledgments
The authors thank the Kavli Institute for Theoretical Physics (KITP), where this research was initiated and partly performed. The KITP is supported, in part, by the National Science Foundation under Grant No. NSF PHY-1748958. S.V. thanks Matthew Fisher for helpful discussions. U.A. thanks Ali Lavasani for helpful discussions. I.L. acknowledges  support from the Gordon and Betty Moore Foundation through Grant GBMF8690 to UCSB. This work was supported by the Simons Collaboration on Ultra-Quantum Matter, which is a grant from the Simons Foundation (651440, U.A.).

\appendix

\bibliography{references}

\appendix
\section{Lattice Partition Function and the Annealed Phase Transition}\label{app:lattice_partition}
\begin{figure}[t]
$\begin{array}{cc}
             \includegraphics[width=.23\textwidth]{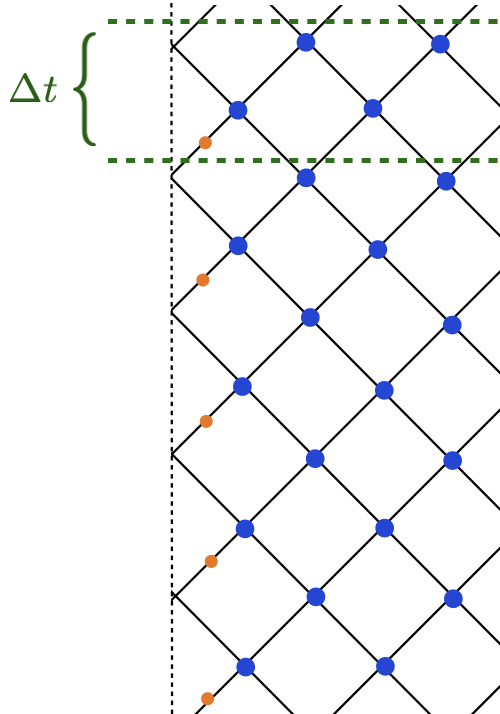}  &
              \includegraphics[width=.22\textwidth]{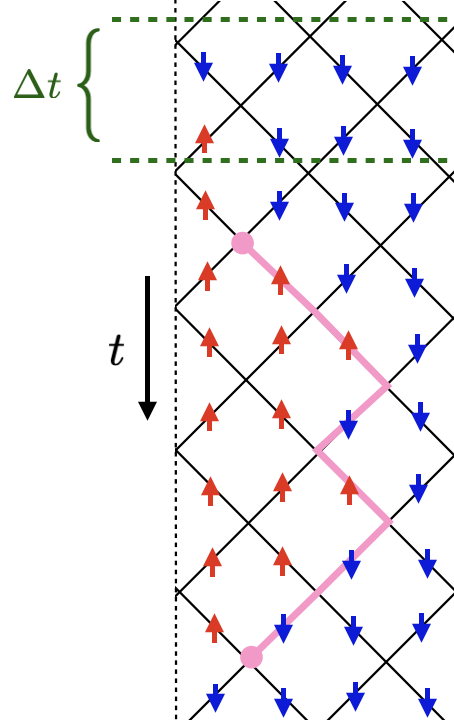}\\
              \text{(a)} & \text{(b)}
\end{array}$
             \caption{A depiction of the quantum circuit which is applied to the qudit chain is shown in (a).  Here, each blue vertex indicates the application of a two-site unitary gate while the orange sites indicate the periodic application of a single-qudit depolarizing channel.  The calculation of the corresponding Ising partition sum can be performed, with spin configurations living on bonds of the square lattice, as in (b), and which are naturally thought of as propagating in the indicated ``time" direction by the transfer matrix for the Ising magnet.  Shown is a contribution to $Z_{f}(t=5)$, where the domain wall is created by the dissipation at the initial time, and is annihilated four timesteps later.  The trajectory of the Ising domain wall can be thought of as a path on the lattice, which starts from the first unitary gate which acts on a pair of anti-aligned spins, and ends when the domain wall is annihilated.}
         \label{fig:DW}
\end{figure}

The annihilation of the Ising domain wall at the boundary, or the free propagation of the domain wall through the bulk describe two distinct phases which may be accessed by tuning the dissipation strength, as described in detail in Sec. \ref{sec:I}.  Here, we make this connection precise by studying the lattice partition function for the domain wall using the weights derived in Sec. \ref{subsec:stat_mech}.  We consider the quantum circuit evolution shown schematically in Fig. \ref{fig:DW}a, where each site (in blue) denotes the action of a two-site unitary gate on a qudit chain, while dissipation (in orange) acts periodically on the boundary qudit.  Let $Z(T)$ denote the partition function for the domain wall propagating for a time $T$, defined so that at the initial and final times, the domain wall is absent (i.e. has been annihilated at the $x=0$ interface).  This partition sum may be calculated as follows.  First, we define $Z_{a}(t)$ to be the partition functions when there is no domain wall for a time interval $t$ (it has been annihilated), while $Z_{f}(t)$ is the partition function when the domain wall is created at the $x=0$ interface, wanders and first returns back to the interface after a time $t$, after which it is then annihilated (the domain wall is free).  With these definitions, we observe that $Z_{0}({T})$ is given by summing over all possible domain wall histories as
\begin{align}\label{eq:series}
&Z(T) = Z_{a}(T) + Z_{f}(T) + \sum_{t < T}Z_{a}(t)Z_{f}(T-t) + \cdots%\nonumber\\
\end{align}
where the ellipsis denotes all possible domain wall configurations in which, at intermediate timesteps, the domain wall wanders away or is annihilated at the interface.

It is convenient to consider the discrete Laplace transform of the partition function
\begin{align}\label{eq:z(w)}
z(w) \equiv \sum_{T\ge 0}w^{T}Z(T).
\end{align}
The inverse of this transformation is given by
\begin{align}
Z(T) = \frac{1}{2\pi i}\oint_{\Gamma} dw \frac{z(w)}{w^{T+1}}
\end{align}
where the contour $\Gamma$ encloses the origin in the complex $w$ plane.  This relation is easily verified by substituting Eq. (\ref{eq:z(w)}).  As a result, the smallest real singularity of $z(w)$ -- denoted $w_{*}$ -- determines the behavior of the partition function at long times.  Equivalently, the free energy density $f = -T^{-1}\log Z$ is given by
\begin{align}
f \overset{T\rightarrow\infty}{\sim} \log w_{*}
\end{align}
The Laplace transform of $Z(T)$ is straightforward to evaluate, since each term in the expansion (\ref{eq:series}) is a discrete convolution of products of $Z_{a}$ and $Z_{f}$.  As a result, the Laplace transform of each term in this sum is simply the product of the Laplace transformations of appropriate products of $Z_{a}$ and $Z_{f}$.  We thus find that
\begin{align}\label{eq:Z_laplace}
z(w) &= \frac{z_{a}(w) + z_{f}(w) + 2\,z_{a}(w)z_{f}(w)}{1 - z_{a}(w)z_{f}(w)}
\end{align}
with $z_{a}(w)$ and $z_{f}(w)$ defined as the Laplace transforms of $Z_{a}(t)$ and $Z_{f}(t)$, respectively.  

Observe that $Z_{a}(t) = (1-p)^{2t}$ so that
\begin{align}
z_{a}(w) = \frac{w(1-p)^{2}}{1-w(1-p)^{2}}
\end{align}
Similarly, we note that when $t \ge 2$
\begin{align}
Z_{f}(t) = \frac{p(2-p)}{q}\left(\frac{q}{q^{2}+1}\right)^{2t-3}N_{2t-4}
\end{align}
Here, $p(2-p)/q$ is the weight to create the Ising domain wall, as indicated in Eq. (\ref{eq:D}).  The domain wall is acted upon by $2t-3$ two-site unitary gates, incurring a weight $q/(q^{2}+1)$ for the action of each gate.  Finally, $N_{2k}$ is the number of walks on the rotated square lattice -- such as the one shown in Fig. \ref{fig:DW}b -- which start at a site closest to the boundary, and which  and return to the same point after $2k$ steps, without touching the boundary.  This counting of paths is easily determined to be
\begin{align}
N_{2k} = \left(\begin{array}{c} 2k\\k\end{array}\right) - \left(\begin{array}{c} 2k\\k+1\end{array}\right).
\end{align}

Performing the Laplace transform thus yields
\begin{align}
z_{f}(w) = \frac{p(2-p)}{2q}\frac{w (q^{2}+1)}{q}\left[1 - \sqrt{1 - \frac{w}{w_{1}(q)}}\right]%\frac{w(1-p^{2})}{2q^{2}}\left[1 + q^{2} - \sqrt{1 + q^{4} + 2q^{2}(1-2w)}\right]\nonumber
\end{align}
which has a singularity when the argument of the square root vanishes at 
\begin{align}
    w_{1}(q) \equiv (q^{2}+1)^{2}/4q^{2}.
\end{align}
We note that $z(w)$ is also singular at $w = w_{2}$ such that $z_{a}(w_{2})z_{f}(w_{2}) = 1$.  Finally, we note that while $z_{a}(w)$ contains a pole at $w = 1/(1-p)^{2}$, it is clear from (\ref{eq:Z_laplace}) that this does not give rise to a singularity in $z(w)$.

When $p > p_{c}$, the smallest real singularity of $z(w)$ occurs at $w = w_{1}(q)$, so that the free energy
\begin{align}
f = 2\log\left(\frac{q^{2}+1}{2q}\right) \hspace{.2in} (p > p_{c})
\end{align}
 A phase transition occurs at $p = p_{c}$ when the two singularities merge $w_{1} = w_{2}$, and for $p < p_{c}$ the singularity at $w_{*} = w_{2}$  determines the free energy density.  The phase transition therefore occurs when 
 \begin{align}
 z_{f}(w_{1})z_{a}(w_{1}) = 1
 \end{align}
This equation may be solved numerically to obtain $p_{c}$ for any finite $q$. The critical probability increases with increasing Hilbert space dimension $q$.  In the limit $q\rightarrow\infty$, we may analytically solve this equation to find that $p_{c}$ approaches one as
\begin{align}
p_{c} = 1 - O(q^{-2})
\end{align}
so that the phase transition is absent when the on-site Hilbert space dimension is strictly infinite.

Finally, we may study the singular part of the free energy near the transition at $p=p_{c}$.  Expanding the equation $z_{f}(w_{2})z_{a}(w_{2}) = 1$ for $p = p_{c} - \delta p$ with $\delta p \ll p_{c}$ yields the result that the singularity $w_{*} = w_{2} = w_{1} - \delta w$ where $\delta w \sim (\delta p)^{2}$.  As a result, the free energy difference vanishes when approaching the critical point as 
\begin{align}
\Delta f(p) \equiv f(p_{c}) - f(p) \overset{p\rightarrow p_{c}^{-}}{\sim} (p - p_{c})^{2}
\end{align}
On general grounds, the singular part of the free energy density should vanish as $\Delta f \sim 1/\xi_{\parallel}$ where $\xi_{\parallel}$ is the correlation length along the time direction.  This correlation length thus diverges as $\xi_{\parallel} \sim (p-p_{c})^{-\nu_{\parallel}}$ with $\nu_{\parallel} = 2$.

Finally, we may determine the typical length of an excursion $\ell_{\perp}$ that the domain wall will make into the bulk of the quantum circuit, and how this distance diverges as we approach the phase transition from the pinned phase $p \le p_{c}$.  First, observe that the weight for the Ising domain wall to make an excursion for a time $t$ is $Z_{f}(t)/Z(t)$.  Then the typical duration of an excursion is 
\begin{align}
\tau = \frac{\displaystyle\sum_{t}t \,Z_{f}(t)/Z(t)}{\displaystyle\sum_{t}Z_{f}(t)/Z(t)} \sim \frac{\displaystyle\sum_{t}t \,w_{*}^{t} Z_{f}(t)}{\displaystyle\sum_{t}w_{*}^{t}Z_{f}(t)} = \frac{\partial\ln Z_{f}(w)}{\partial\ln w}\Big|_{w = w_{*}}\nonumber
\end{align}
where in the second expression, we have used the fact that $Z(t) \overset{t\rightarrow\infty}{\sim} w_{*}^{-t}$.  On approaching the transition from the localized phase $p = p_{c} - \delta p$, the singularity $w_{*} = w_{2} = w_{1} - \delta w$ with $\delta w \sim \delta p^{2}$, as derived previously, which yields the result that $\tau \sim (p_{c} - p)^{-1}$ as $p\rightarrow p_{c}^{-}$. Assuming a diffusive wandering of the domain wall, the transverse distance covered by the domain wall diverges on approaching the depinned phase as
\begin{align}
\ell_{\perp} \overset{p\rightarrow p_{c}^{-}}{\sim} (p_{c} - p)^{-1/2}
\end{align}

Approaching the phase transition, when $\ell_{\perp}\gg x_{0}$, the probability that the domain wall has reached a point $y\ge x_{0}$ is approximately $P(x_{0},t) = 1 - O(x_{0}/\ell_{\perp})$.  Substituting this into Eq. (\ref{eq:annealed_MI}) yields the result that the annealed mutual information vanishes as $I^{(\mathrm{ann})}_{A,R} \sim \ell_{\perp}^{-1} \sim (p_{c}-p)^{\beta}$ (with $\beta \equiv 1/2$) when approaching the phase transition.  This behavior, along with the knowledge of $\nu_{\parallel} = 2$ motivates the finite-size scaling form for the annealed mutual information, which we use in the main text $I^{(\mathrm{ann})}_{A,R}(T) = T^{-\beta/\nu}F(T^{1/\nu}(p-p_{c}))$.

\section{Alternative random circuit protocols}\label{app:protocols}

To show that the phase transition in the mutual information persists irrespective of the precise form of the boundary dissipation and scrambling dynamics, here we introduce and examine four different protocols for the random circuit. We consider the following two types of time evolution, each of them with two different realizations of the boundary dissipation.

$\bullet$  Random boundary dissipation + maximal Clifford scrambling. In each time step, the dissipation acts on  the leftmost qubit with probability $p$. Scrambling is provided by random Clifford gates arranged in a brickwork structure. Therefore, the relative strength of the dissipation compared to the efficiency of scrambling is tuned through the parameter $p$.

$\bullet$  Periodic boundary dissipation + sparse Clifford scrambling. The dissipation acts on  the leftmost qubit periodically, with periodicity $T_{\rm period}$. The unitary gates providing the scrambling of information are applied in a sparse brickwork structure, where each gate in the brickwork is a random Clifford unitary with probability $p_U$, and the identity with probability $1-p_U$. In this scenario, the relative strength of the dissipation compared to the efficiency of scrambling is determined by two parameters, $T_{\rm period}$ and $p_U$.

\begin{figure*}[t]
    \includegraphics[width=.38\linewidth]{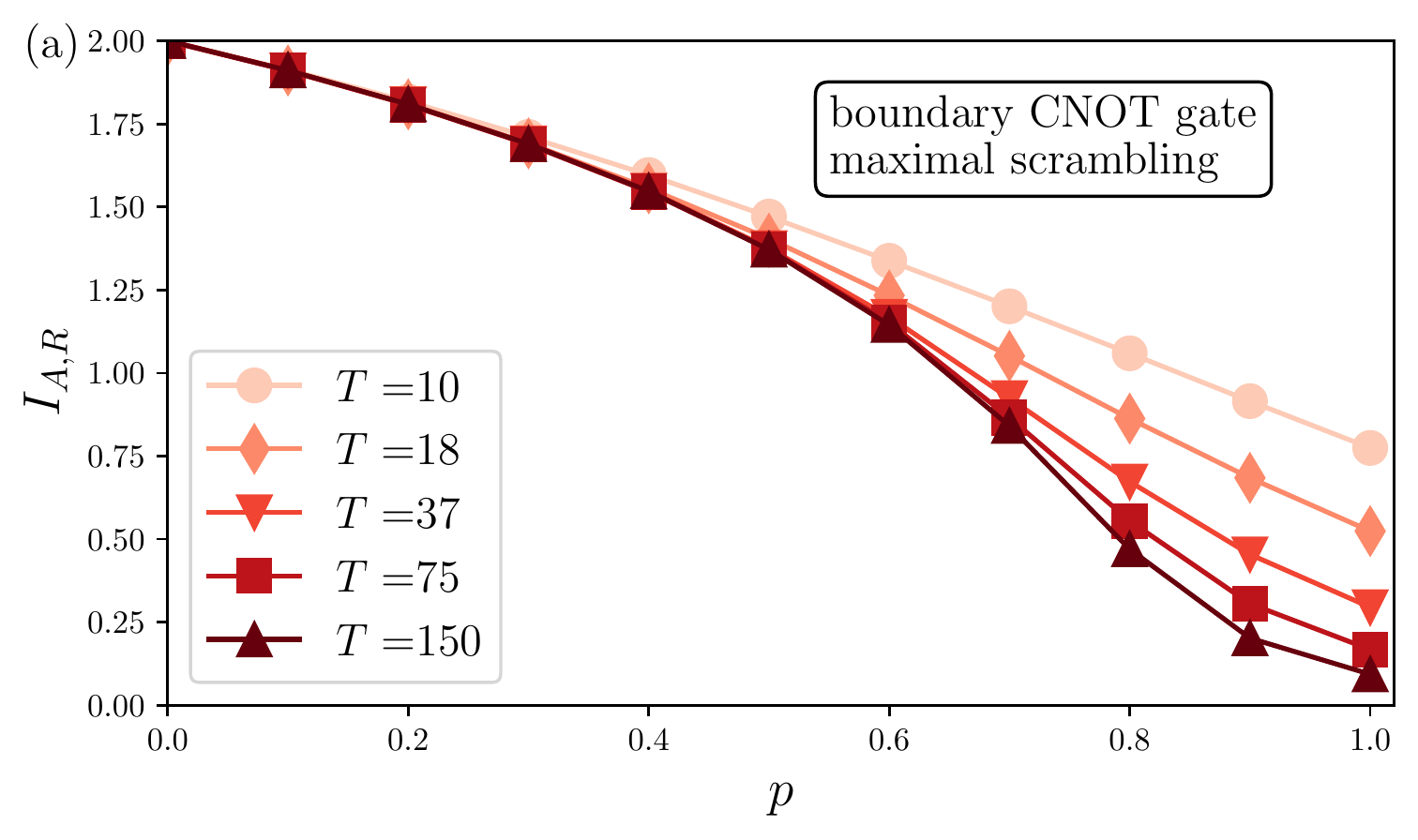}
    \includegraphics[width=.38\linewidth]{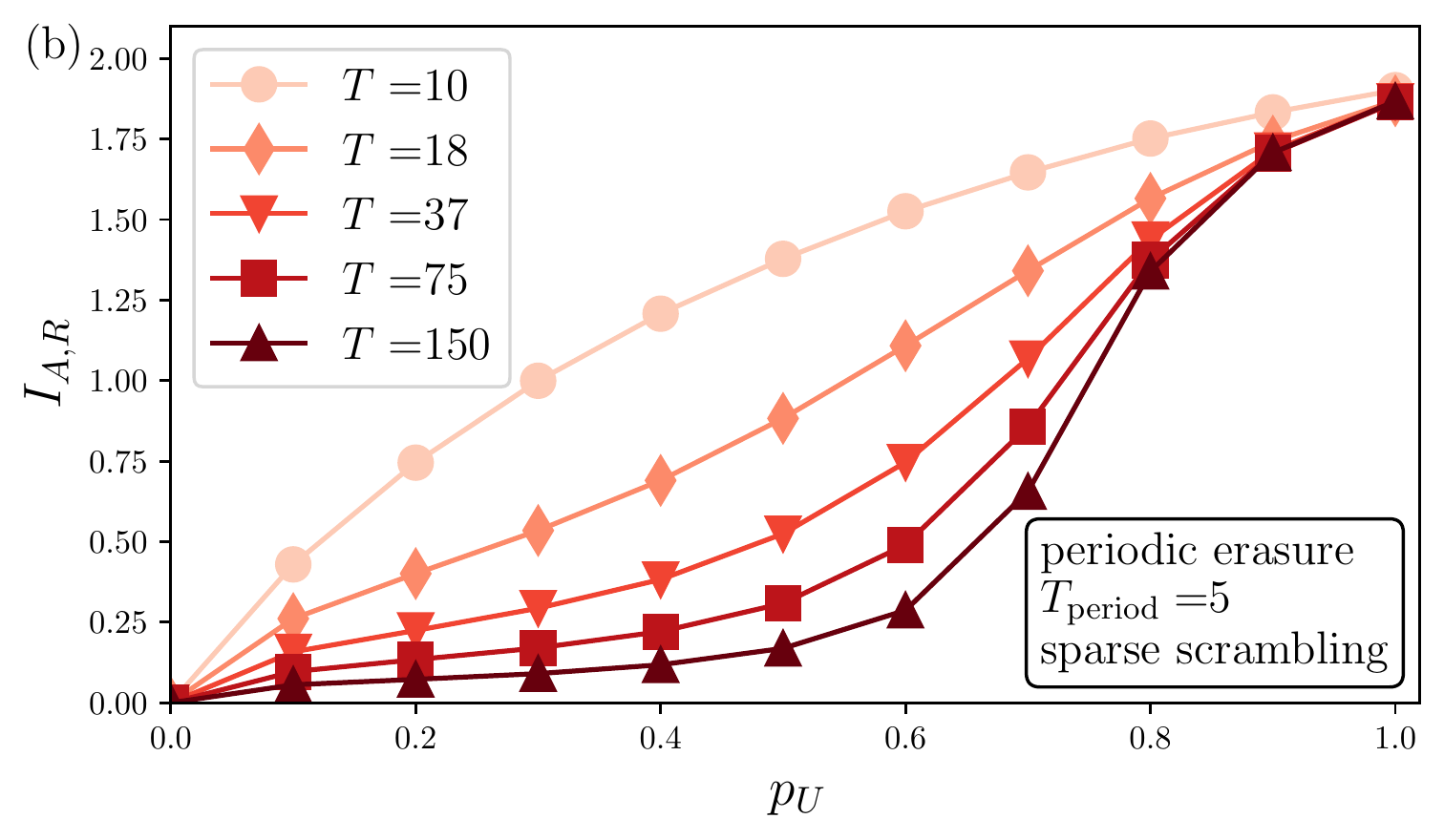}
    \includegraphics[width=.38\linewidth]{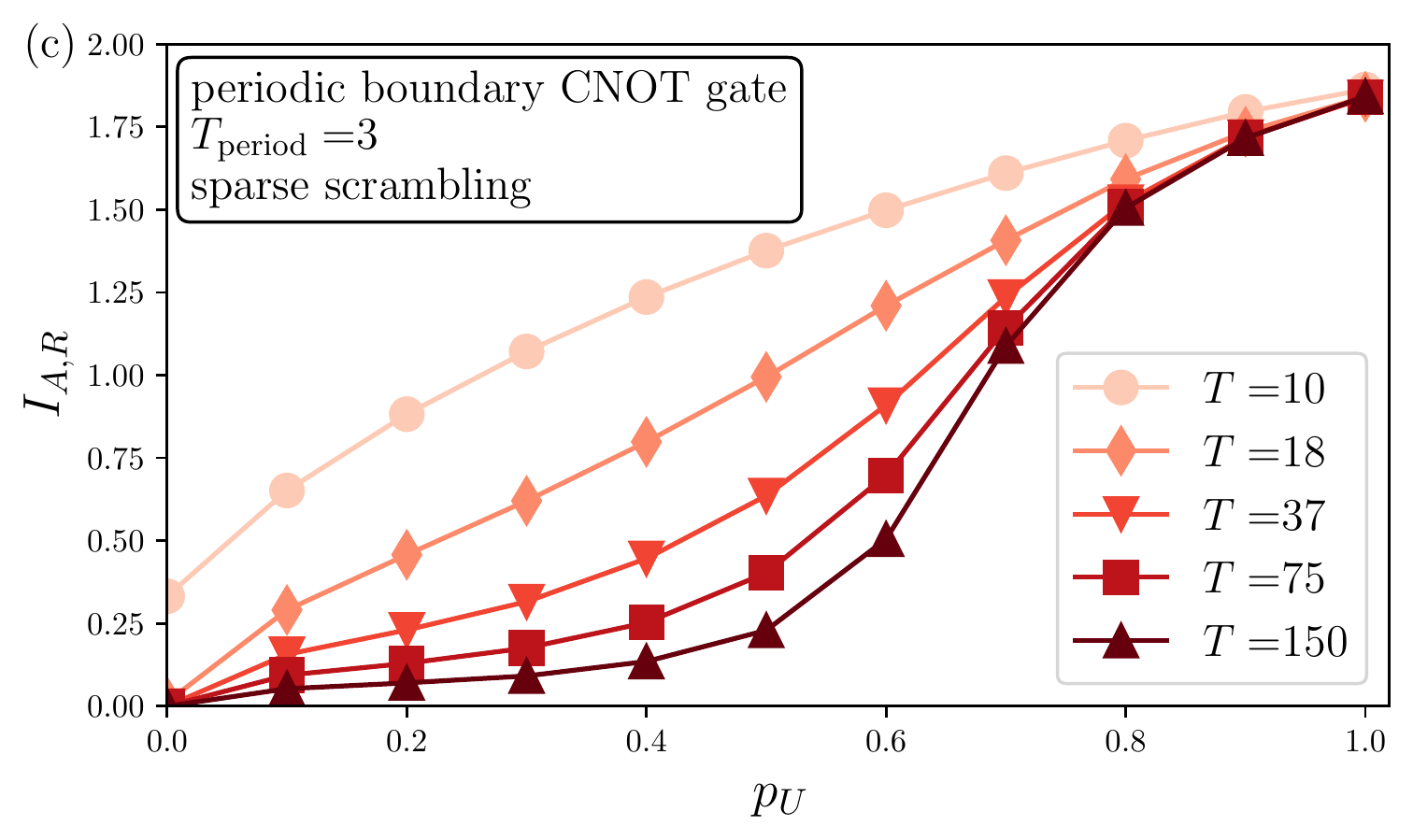}
   \caption{Coding transition induced by a single boundary for different circuit protocols. Mutual information between the input and the output of the circuit (a) as a function of dissipation strength $p$ for boundary dissipation realized as a CNOT coupling to an ancilla qubit with maximal bulk Clifford scrambling, (b)-(c) varying the strength of sparse bulk Clifford scrambling $p_U$ with periodic boundary erasure channel (b), or periodic boundary CNOT gate to an ancilla (c). All data are consistent with a coding transition between a phase with partially protected information for weak dissipation / strong enough scrambling, and a dissipative phase with all encoded information lost. No pre-scrambling step was used for these plots.}
    \label{fig:otherprotocol}
\end{figure*}

As described in the main text, the Bell pair is encoded in the initial state at the left boundary, optionally followed by a pre-scrambling step logarithmic or linear in system size, depending on the type of phase transition that we consider. We note that the pre-scrambling is realized by a full or sparse brickwork of Clifford unitary gates, in the first and second types of dynamics, respectively.  

As mentioned above, we consider two different realizations of the boundary dissipation.

$\bullet$  Boundary erasure channel. The dissipation acts by deleting the information stored in the leftmost qubit.

$\bullet$  Coupling to an ancilla qubit. Here, we first couple the leftmost qubit of the system to an ancilla qubit through a CNOT gate, and then trace out the ancilla. In the stabilizer formalism, this operation results in deleting all stabilizers containing a $Y$ or $Z$ Pauli operator at the left end of the chain. To restore rotational invariance and obtain a smooth limit $p_U\rightarrow 0$, for sparse Clifford scrambling we also act with a random single site Clifford gate on the leftmost qubit before applying the CNOT gate.

In the main text we mainly focused on the case of random boundary dissipation and maximal Clifford scrambling, with the dissipation realized as a boundary erasure channel. We also briefly commented on the effect of a periodic boundary noise, modifying the critical properties for linear pre-scrambling compared to the random case. Below we provide supplementary numerical results for the other protcols, showing a similar phase transition in the mutual information.

We show the coding transition in the mutual information without pre-scrambling, induced by a single boundary with aspect ratio $T/L<1$, in Fig.~\ref{fig:otherprotocol} for three different protocols.  We cross the phase transition by tuning the strength of dissipation $p$ in Fig.~\ref{fig:otherprotocol}a, realized  with a random CNOT coupling between the boundary spin and an ancilla qubit. In contrast, in Fig.~\ref{fig:otherprotocol}b and c the tuning parameter is the strength of sparse bulk scrambling $p_U$,  while we  apply a fixed strength periodic boundary dissipation,  realized as an erasure channel in Fig.~\ref{fig:otherprotocol}b, and as a CNOT gate with an ancilla qubit in Fig.~\ref{fig:otherprotocol}c. We recover the coding transition between a phase with partially protected coherent information and a phase where all information is destroyed for all protocols. Due to the difficulties in fitting critical exponents from finite size data mentioned in the main text, we leave the detailed study of critical properties for future work. In the cases with periodic boundary dissipation we used $T_{\rm period}=5$ (b), and $T_{\rm period}=3$ (c).

\end{document}